\documentclass[useAMS,usenatbib]{mn2e}
\usepackage{epsfig}
\usepackage{amssymb}
\usepackage{amsmath}
\usepackage{natbib}
\usepackage{threeparttable} 
\usepackage{graphicx}
\usepackage{setspace}
\usepackage{longtable}
\usepackage{multicol}
\usepackage{subfigure}
\usepackage{verbatim}
\newcommand\apj{ApJ}
\newcommand\apjl{ApJ}
\newcommand\apjs{ApJS}
\newcommand\aap{A\&A}
\newcommand\aapr{A\&A~Rev.}
\newcommand\mnras{MNRAS}
\newcommand\iaucirc{IAU~Circ.}
\newcommand\gca{Geochim.~Cosmochim.~Acta}
\def\hide#1{}

\def\mathnew{\mathsurround=0pt}
\def\simov#1#2{\lower .5pt\vbox{\baselineskip0pt \lineskip-.5pt
        \ialign{$\mathnew#1\hfil##\hfil$\crcr#2\crcr\sim\crcr}}}

\def\simless{\mathrel{\mathpalette\simov <}}

\title[Source spectrum and kHz QPOs in 4U 1636$-$53]{Relation between spectral 
changes and the presence of the lower kHz QPO in the neutron-star low-mass 
X-ray binary 4U 1636$-$53}

\author[G. Zhang et al.]{Guobao Zhang$^{1}$\thanks{E-mail: guobao.zhang@nyu.edu},  Mariano M\'endez$^{2}$,
Andrea Sanna$^{3}$,  Evandro M. Ribeiro$^{2}$ 
\newauthor
and Joseph D. Gelfand$^{1,	4}$\\
$^{1}$New York University Abu Dhabi, P.O. Box 129188, Abu Dhabi, United Arab Emirates\\
$^{2}$Kapteyn Astronomical Institute, University of Groningen, P.O. BOX 800, 9700 AV 
Groningen, The Netherlands\\
$^{3}$Dipartimento di Fisica, Universit\'a degli Studi di Cagliari, SP Monserrato-Sestu 
km 0.7, I-09042, Monserrato, Italy\\ 
$^{4}$Center for Cosmology and Particle Physics, New York University, Meyer Hall of Physics, 
4 Washington Place, New York, NY 10003
}

\begin{document}


\maketitle

\label{firstpage}

\date{Accepted. Received; in original form}

\begin{abstract}

We fitted the $3-180$-keV spectrum of all the observations of the neutron-star 
low-mass X-ray binary 4U 1636$-$53 taken with the {\it Rossi X-ray Timing Explorer} 
using a model that includes a thermal Comptonisation component. We found that in 
the low-hard state the power-law index of this component, $\Gamma$,  gradually
increases as the source moves in the colour-colour diagram. When the source 
undergoes a transition from the hard to the soft state $\Gamma$ drops abruptly; 
once the source is in the soft state $\Gamma$ increases again and then decreases 
gradually as the source spectrum softens further. The changes in $\Gamma$, together 
with changes of the electron temperature, reflect changes of the optical depth 
in the corona. The lower kilohertz quasi-periodic oscillation (kHz QPO) in this 
source appears only in observations during the transition from the hard to the 
soft state, when the optical depth of the corona is high and changes 
depends strongly upon the position of the source in the colour-colour diagram.
Our results are consistent with a scenario in which the lower kHz QPO reflects 
a global mode in the system that results from the resonance between, the disc 
and/or the neutron-star surface, and the Comptonising corona.

\end{abstract}


\section{introduction}
\label{introduction}

Energy spectra and colour-colour diagrams (CD) are often used to study 
neutron-star low-mass X-ray binaries  
\citep[NS-LMXBs; e.g., ][]{Hasinger89}. The evolution of the energy spectrum and the tracks on 
the CD are thought to be driven by variations of mass accretion rate, and reflect 
changes in the configuration of the accretion flow \citep[e.g., ][]{Hasinger89, Mendez99, Done07}. 
In the low-hard state the accretion rate is low, the disc is truncated at large radii 
\citep[][see however Lin et al. 2007]{GierlinskiDone02, Sanna13, Plant15} and the energy 
spectrum is dominated by a hard/Comptonised power-law component. When the source accretion 
rate increases, the disc truncation radius decreases and eventually reaches the last stable orbit. 
In the high-soft state the accretion rate is high and the energy spectrum is dominated 
by a soft component, possibly a combination of the accretion disc and the neutron star.

The characteristic frequencies (e.g., quasi-periodic oscillations, QPOs) in the power 
density spectra (PDS) of these systems also change with the source luminosity and 
inferred mass accretion rate \citep[e.g., ][]{Mendez99, Belloni05}.  
Kilohertz (kHz) QPOs have been detected in 
many NS-LMXBs \citep[for a review see][and references therein]{Klis06}. The upper 
kHz QPO (from the pair of QPOs the one at the highest frequency) in these systems 
has been interpreted  in terms of characteristic frequencies (e.g., the Keplerian 
frequency) in a geometrically thin accretion disc \citep{Miller98, Stella98}. 
In this scenario, changes of the upper kHz QPO frequency reflect changes of the inner 
disc radius, driven by mass accretion rate. Indeed, the frequency of the 
upper kHz QPOs is strongly correlated with the hard colour  of the source
\citep{Mendez99, Belloni05, Belloni07, Sanna12}.

Several models have been proposed to explain the lower kHz QPO in these systems. 
\cite{Stella98} suggested  a  Lense-Thirring precession model, in which 
the frequencies of the QPOs are associated with the fundamental frequencies 
of geodesic motion of clumps of gas around the compact object. In the 
relativistic resonance model \citep{Kluzniak01, Lee01, Kluzniak04}, the 
kHz QPOs appear at frequencies that correspond to a coupling between two 
oscillations modes of the accretion disc. In the beat-frequency model \citep{Miller98},  
the lower kHz QPO originates from the interaction between the spin frequency of the NS 
and material orbiting at the inner edge of the accretion disc. 
None of these models, however, have so far been able to fully explain all the properties 
of kHz QPOs \citep[e.g., ][]{Jonker02, Belloni05, Altamirano12}.

4U 1636--53 is a NS-LMXB that shows regular state transitions with a cycle of $\sim$40 days 
\cite[e.g., ][]{Belloni07}, making it an excellent source to study 
correlations between its spectral and timing properties. The full range of spectral 
states (low/hard state, high/soft state, transitional state) has been observed in 
this source \citep{Belloni07, Altamirano08}. A pair of kHz QPOs were 
discovered by \cite{Wijnands97} and \cite{Zhang97}. The upper kHz QPO has been 
observed in different states. Its central frequency shows a clear correlation with 
the hard colour of the source \citep{Belloni07, Sanna12}.  The lower kHz-QPO in 
4U 1636--52 is only detected over a narrow range of hard  colour values \citep{Belloni07, Sanna12}. 
The emission mechanism of the lower kHz-QPO is still unclear 
\citep[e.g., ][]{Berger96, Mendez01, Mendez06}

We analysed the broadband energy spectra of 4U 1636--53  
to investigate the evolution of the different spectral and timing
components as a function of the spectral state of the source. A comparison 
the different continuum components in the energy spectrum with the properties of 
the kHz QPOs at the same state may provide an important clue to 
understand the origin of the kHz QPOs and the evolution of the 
accretion flow geometry. In  \S\ref{data} we describe the observations, 
data reduction and analysis methods,  and in \S\ref{result} we present 
the results on the temporal and spectral analysis of these data. 
Finally, in \S\ref{discussion} we discuss our findings and 
summarise our conclusions.

\section{Observations and data analysis}
\label{data}

\subsection{Data reduction}

We analysed the whole archival data (1576 observations) from the Rossi X-ray 
Timing Explorer ({\it RXTE}) Proportional Counter Array  \citep[PCA;][]{Jahoda06} 
and the High-Energy X-ray Timing Experiment \citep[HEXTE; ][]{Rothschild98} 
of the NS-LMXB 4U 1636--53. We reduced the data using the {\sc heasoft} package version 6.13.
We extracted PCA spectra from the Proportional Counter Unit number 2  (PCU-2) only, since 
this was the best-calibrated  detector and the only one which was always on in all the 
observations. To extract the spectra of the source we first examined 
the light curves to identify and remove X-ray bursts from the data. 
For the HEXTE data we generated the spectra using cluster B only, since after 
January 2006 cluster A stopped rocking and could no longer measure the background. 
For each observation we extracted one PCA and HEXTE X-ray spectrum, respectively.
The PCA and HEXTE background spectra were extracted using the standard RXTE tools 
{\sc pcabackest} and {\sc hxtback}, respectively. We built instrument response files 
for the PCA and HEXTE data using {\sc pcarsp} and {\sc hxtrsp}, respectively.

\subsection{Timing analysis}
\label{Timing}

For each observation we computed Fourier power density spectra (PDS) in the 
$2-60$ keV band every 16 s from event-mode data. For this we binned the 
light curves to 1/4096 s, corresponding to a Nyquist frequency of 2048 Hz. 
Before computing the Fourier transform we removed detector dropouts, but we 
did not subtract the background or applied any dead-time correction before 
calculating the PDS. Finally we calculated an average PDS per observation 
normalised as in \cite{Leahy83}. We finally used the procedures described in 
\cite{Sanna12} to detect and fit the QPOs in each PDS. We detected kHz 
QPOs in 581 out of 1576 observations. We detected the lower kHz QPO in 403 
out of those 583 observations.

\subsection{Spectral analysis}
\label{Spectral}

\begin{figure}
    \centering
        \includegraphics[width=3.30in,angle=0]{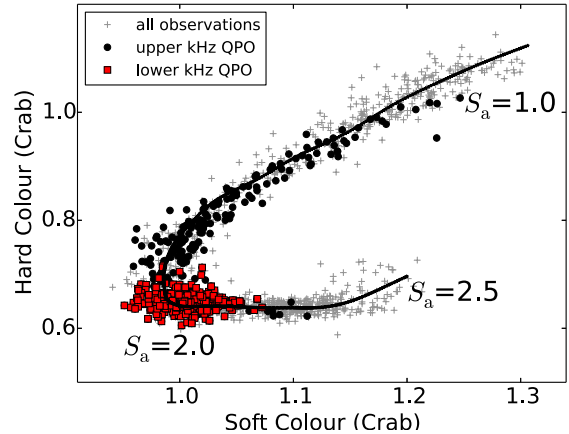}
    \caption{The colour-colour diagram of 4U 1636--53. Each point in the plot corresponds 
    to a single RXTE observation. The filled circles (black) 
    and the filled squares (red) represent the observations with upper and lower kHz 
    QPOs, respectively.   The gray crosses represent observations in which no kHz QPO 
    was detected. The position of the source on the diagram is parametrized by the 
    length of the black solid curve $S_{\rm a}$
             }
    \label{fig:ccd}
\end{figure}

We used the Standard-2 data (16-s time-resolution and 129 channels covering the full
$2-60$ keV PCA band) to calculate X-ray colours of the source \citep[see][for details]{Zhanggb09}. 
We defined hard and soft colours as the $9.7-16.0/6.0-9.7$ keV and $3.5-6.0/2.0-3.5$ 
keV count rate ratios, respectively. 
We show the CD of all observations of 4U 1636$-$53 in Figure \ref{fig:ccd}, with one 
point per RXTE observation. In that figure we parameterised the position of the source in the 
CD by the length of the solid curve $S_{\rm a}$ \citep[see, e.g. ][]{Mendez99}, 
fixing the values of $S_{\rm a}=1$ and $S_{\rm a}=2$ at the top-right and the bottom-left 
vertex of the CD, respectively.

For the spectral analysis of 4U 1636--53, we used the package  {\sc xspec} v12.7 \citep{Arnaud96}.  
We fitted the PCA and HEXTE spectra simultaneously in the 3.0$-$25.0 
and 25.0$-$180.0 keV range, respectively. We included the effect of interstellar absorption 
using the component {\sc phabs} with cross-sections of \cite{Balucinska92} and solar abundances 
from \cite{Anders89}, and we fixed the  column density to $N_{\rm H} = 3.1\times10^{21}$ cm$^{-2}$ \citep{Sanna13}.  
We added a multiplicative factor to the model to account for calibration uncertainties between 
the PCA and HEXTE. We set this factor to unity for the  PCA and left it free for the HEXTE 
spectra.

Many models have been proposed to fit the spectra of accreting NS-LMXBs in the past
\citep[e.g. ][]{Barret01, Lin07}. Most models include at least two components: 
a soft thermal component that represents the emission from the NS surface (or boundary layer) 
and the accretion disc,  and a hard Comptonised component that represents the emission from 
a corona of hot electrons.

In the spectral fitting of 4U 1636--53 we used the same continuum model as in \cite{Sanna13},
who fit {\it XMM-Newton} data down to 0.8 keV.  We used a multi-colour disc blackbody 
({\sc diskbb} in {\sc xspec}) to fit the thermal emission from the disc, a single-temperature black body 
({\sc BB}, {\sc bbodyrad} in {\sc xspec}) to fit the thermal emission from the NS surface (or boundary 
layer), and a Comptonisation model, {\sc nthcomp}, to fit the Comptonised component
\citep{Zdziarski1996, Zycki1999}.

The parameters of {\sc bbodyrad} and {\sc diskbb} are the blackbody temperature, $kT_{\rm bb}$, 
the temperature at the inner disc radius, $kT_{\rm dbb}$, and their normalisations, respectively. 
The parameters of {\sc nthcomp} are the asymptotic power-law  photon index, $\Gamma$,  
the electron temperature of the corona, $kT_{\rm e}$,  the seed 
photon temperature, $kT_{\rm seed}$,  and the normalisation. The seed photons can be from either the
NS surface and boundary layer  or the accretion disc. \cite{Sanna13} analysed the spectra of 
six observations of 4U 1636--53 taken with {\it XMM-Newton} and {\it RXTE} simultaneously. 
They tried both the {\sc bbodyrad} or the {\sc diskbb} component as the source of seed photons for 
{\sc nthcomp}, and concluded that {\sc diskbb} was the best option. Therefore, in 
this work we chose {\sc diskbb} as the source of the soft seed photons and we linked the seed 
photons temperature in {\sc nthcomp} to the temperature of the {\sc diskbb} 
component. In the fits, the value of $\Gamma$ was constrained to be larger or equal 
to 1.1 \citep{GierlinskiDone02, Sanna13} and the value of $kT_{\rm e}$ was restricted 
to be larger or equal to 2.5 keV \citep{GierlinskiDone02, Lyu14}

From the initial fits, there were always relatively broad residuals between 6 and 7 keV 
in the {\it RXTE}/PCA spectra. These residuals likely reflect the presence of an iron line 
in the data. The iron line has also been confirmed with {\it XMM-Newton} and {\it Suzaku} observations 
\citep{Sanna13, Lyu14}. We then added a Gaussian component with a variable width to the 
spectrum to fit this line. Due to the low spectral resolution of the PCA ($\sim$1 keV at $6$ 
keV), we can not constrain the central value of the line very well. So we fixed the line at 
6.5 keV in all our fittings.

Fitting simultaneously data taken from {\it XMM-Newton} and {\it RXTE}, \cite{Sanna13} found that 
the temperature of {\sc bbodyrad} was $kT_{\rm bb}\sim1.7$ keV, 
and remained more or less constant as the source moved across the CD. 
They also found that the temperature at the inner edge of the accretion disc 
ranged from $\sim$0.2 keV to $\sim$0.8 keV when the source moved from the low/hard 
to the high/soft state. Since the PCA only extends down to 3 keV, $kT_{\rm bb}$ and 
$kT_{\rm dbb}$ are difficult to constrain simultaneously. Therefore, we initially left 
$kT_{\rm dbb}$ free and, following the results of \cite{Sanna13}, we fixed $kT_{bb}$ 
to 1.7 keV. We then repeated the analysis fixing $kT_{\rm bb}$ to each of these values: 
1.5, 1.6, 1.8, 1.9 and 2.0 keV.

From the previous analysis we found that $kT_{\rm dbb}$, linked to  $kT_{\rm seed}$  in {\sc nthcomp},  
was not well constrained during the fits. We therefore interpolated the value of $kT_{\rm dbb}$ along 
the CD using the {\it XMM-Newton--RXTE} spectra fitting results in \cite{Sanna13}. 
We divided the data into 7 groups based on their $S_{\rm a}$ values: $<1.2$, $1.2-1.4$, 
$1.4-1.6$, $1.6-1.8$, $1.8-2.0$, $2.0-2.2$ and $> 2.2$. In each group we fixed  
$kT_{\rm dbb}$ to the interpolated temperatures for $S_{\rm a}$ equal to 1.1, 1.3, 1.5, 
1.7, 1.9, 2.1, and 2.35, respectively. In this case we left $kT_{\rm bb}$ free during the fits.

\section{Results}
\label{result}

\subsection{X-ray spectral evolution}
\label{xray_spectral}

\begin{figure*}
\centering 
\resizebox{1.\columnwidth}{!}{\rotatebox{0}{\includegraphics{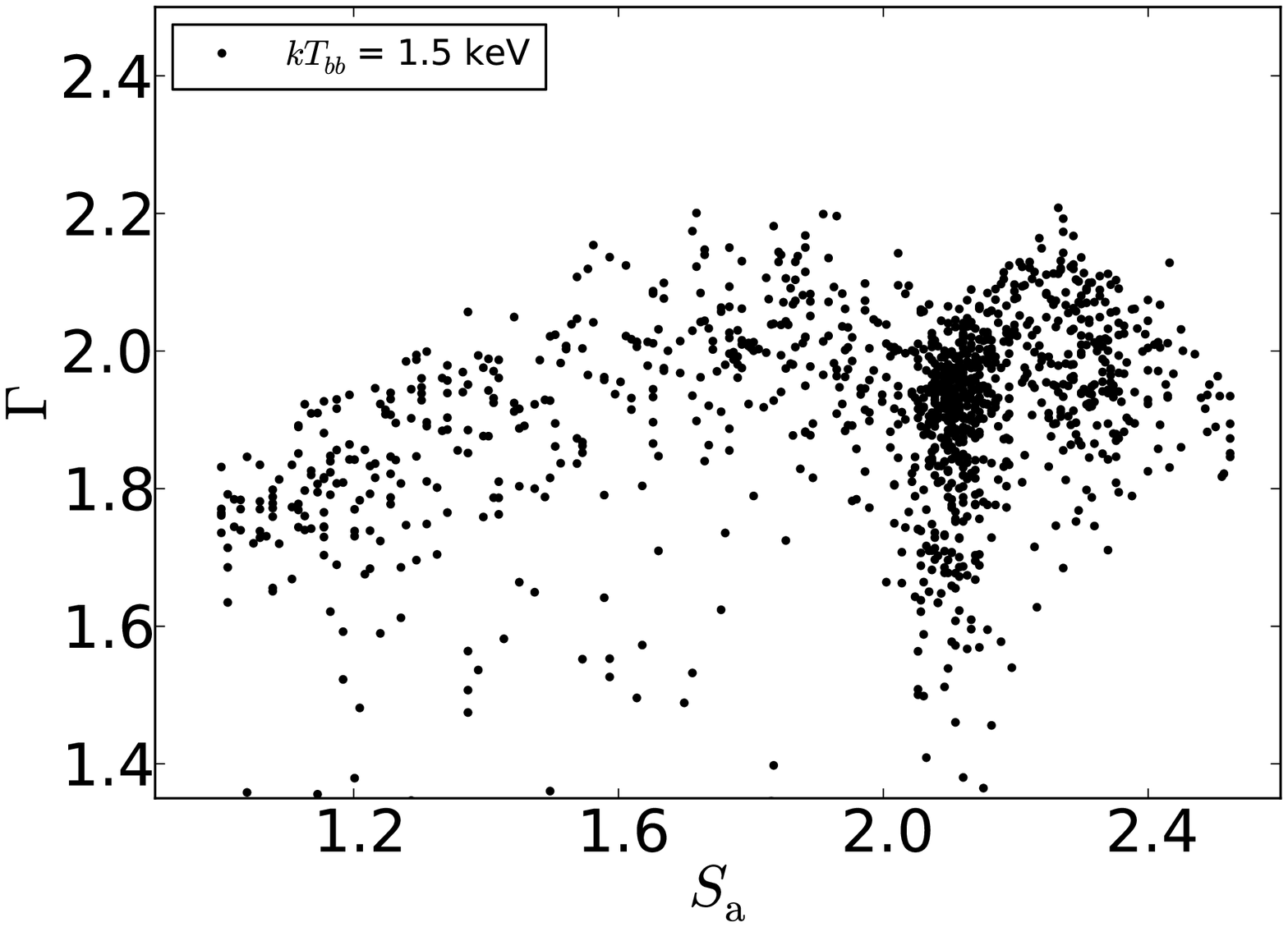}}}           
\resizebox{1.\columnwidth}{!}{\rotatebox{0}{\includegraphics{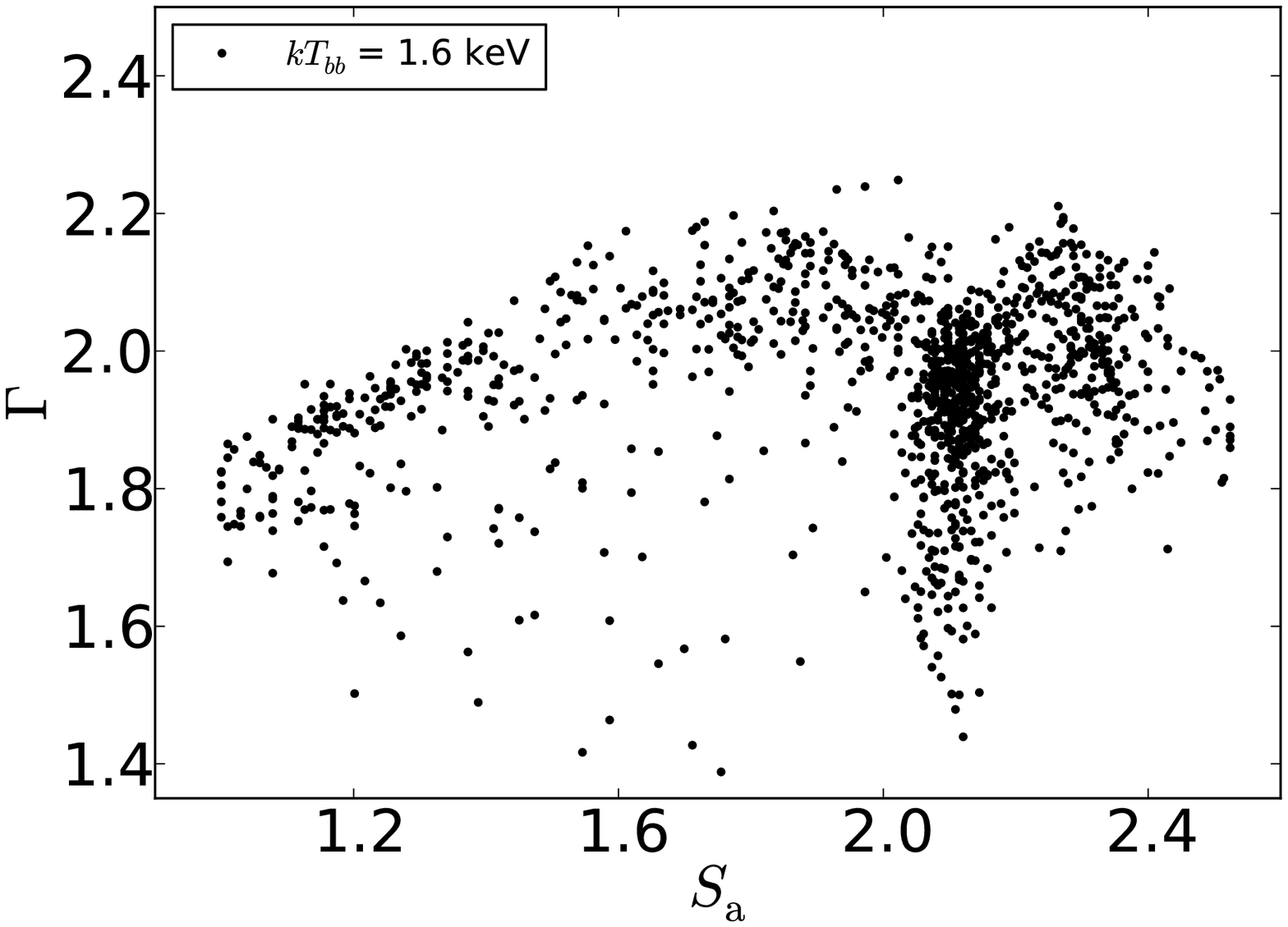}}}           
\resizebox{1.\columnwidth}{!}{\rotatebox{0}{\includegraphics{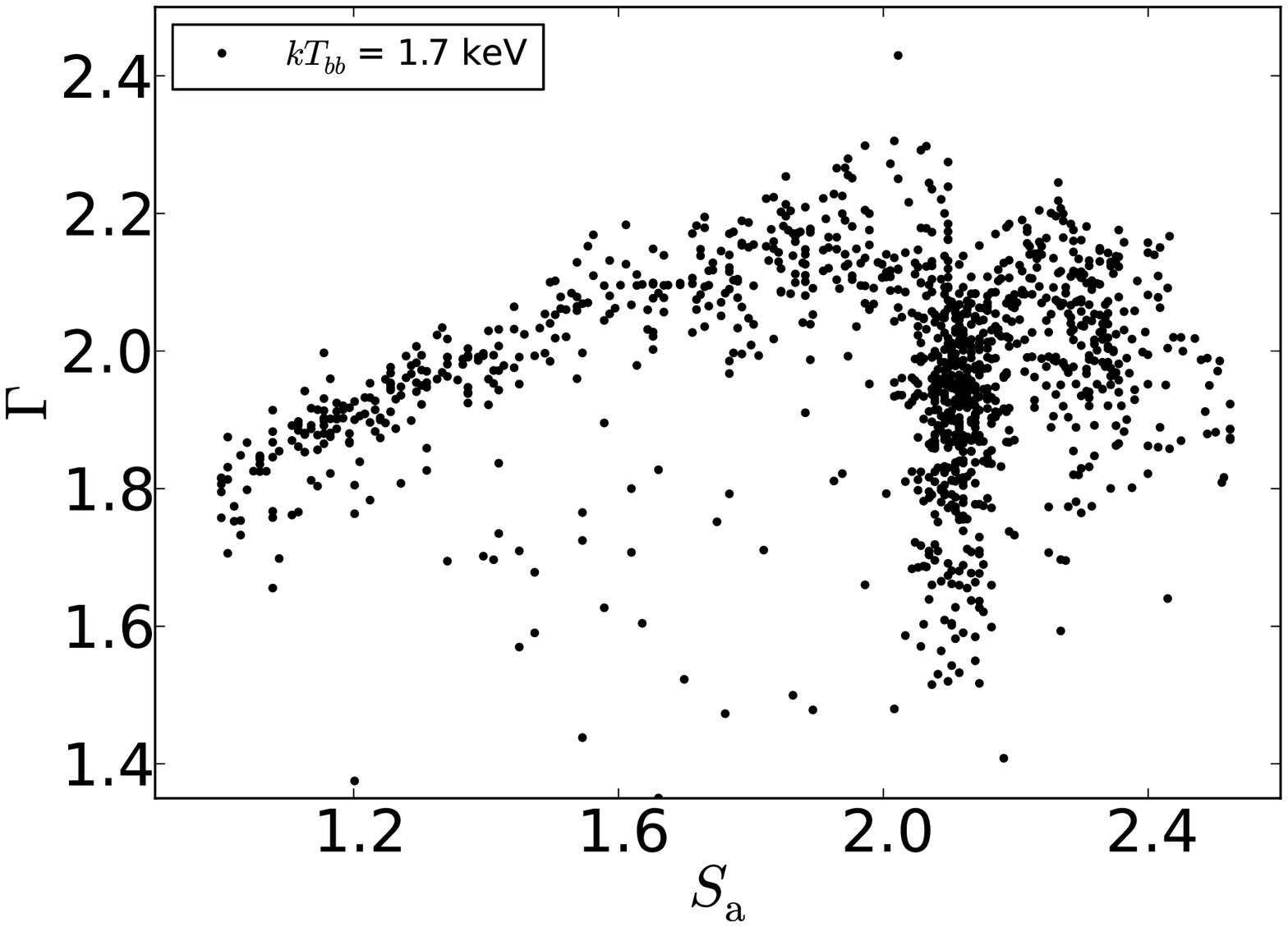}}}           
\resizebox{1.\columnwidth}{!}{\rotatebox{0}{\includegraphics{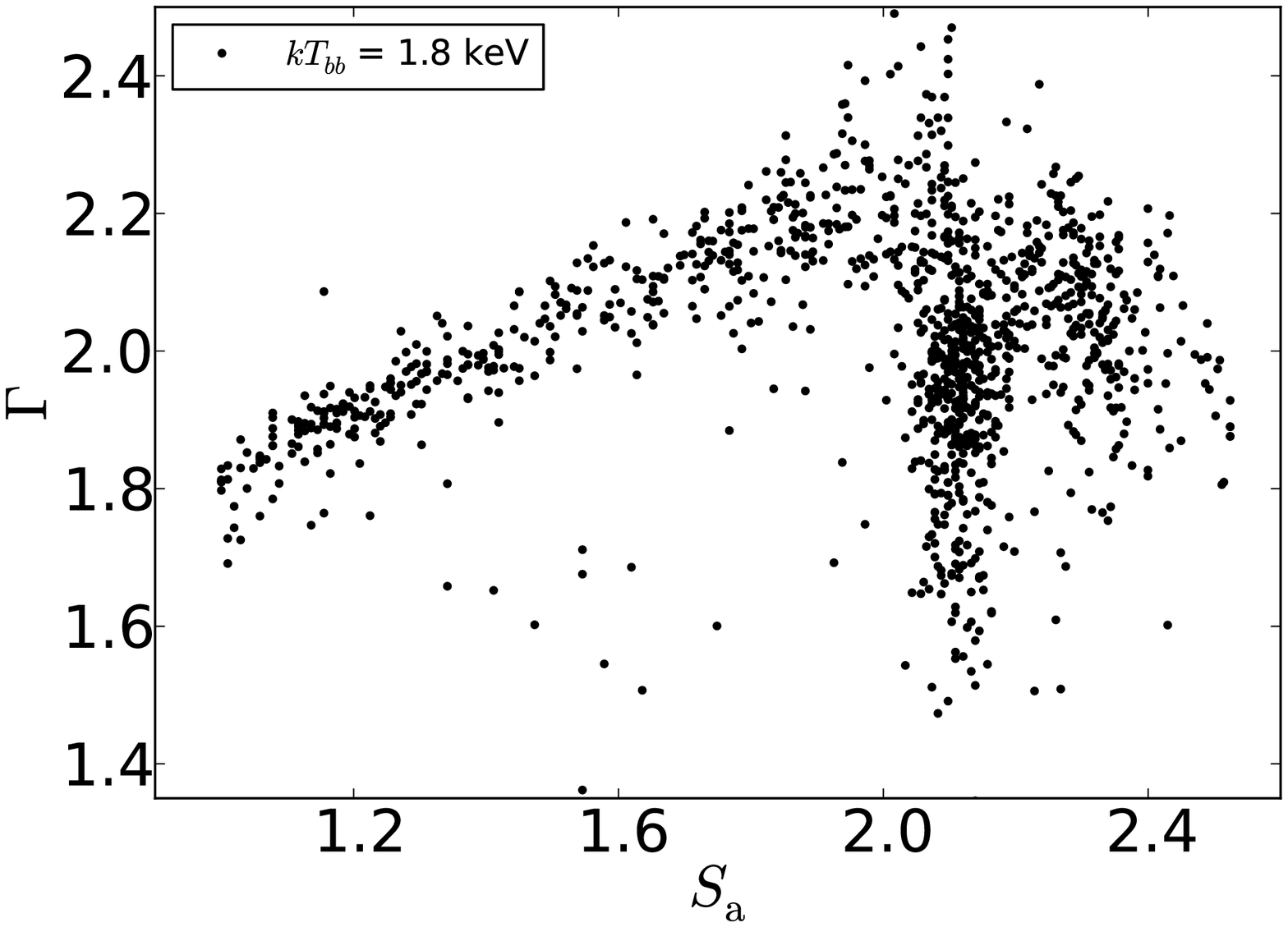}}}            
\resizebox{1.\columnwidth}{!}{\rotatebox{0}{\includegraphics{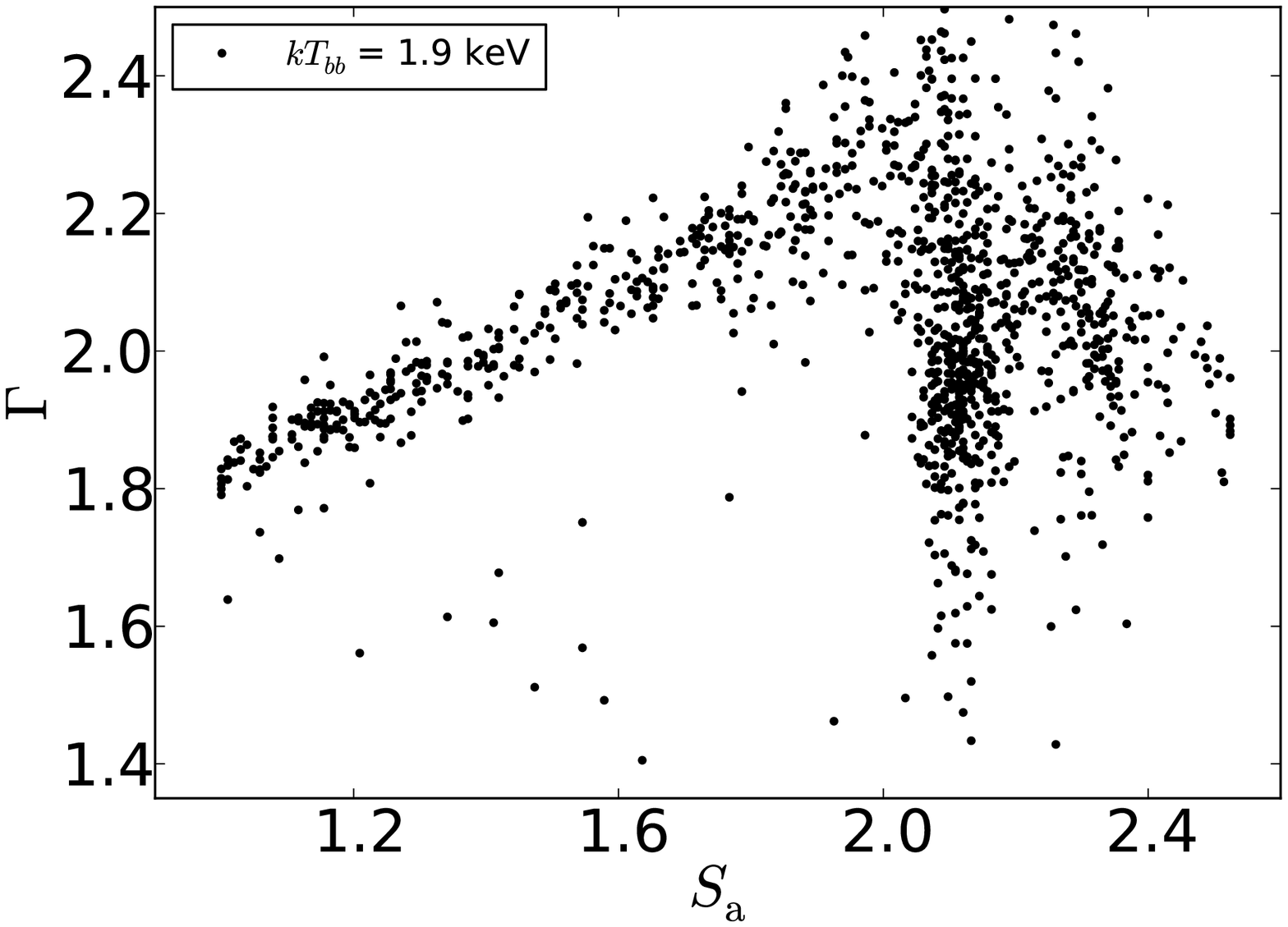}}}           
\resizebox{1.\columnwidth}{!}{\rotatebox{0}{\includegraphics{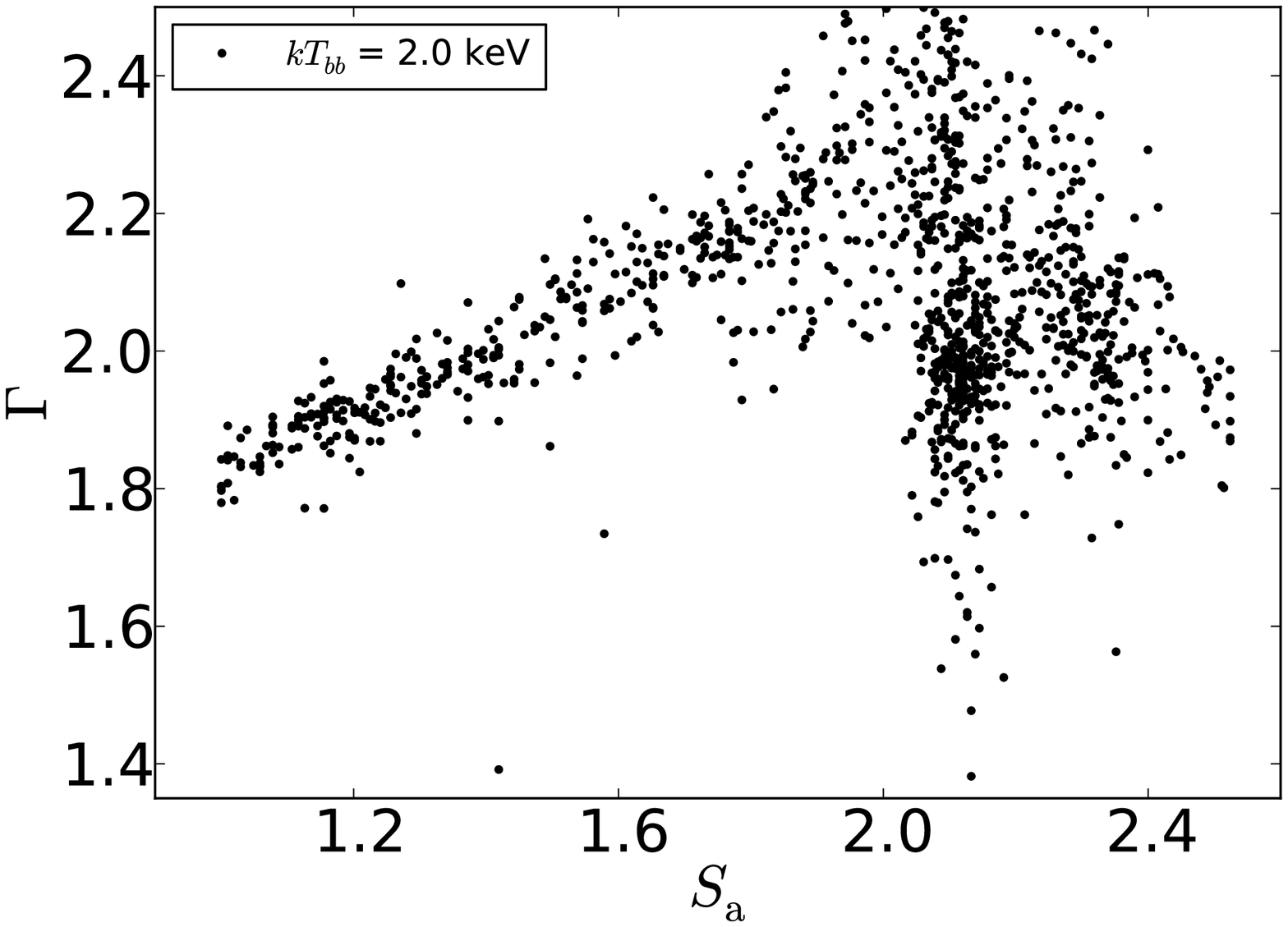}}}             	                
\caption{The power-law index, $\Gamma$, of the {\sc nthcomp} component as a   
function of $S_{\rm a}$ in 4U 1636--53. The temperature of the blackbody component, $kT_{\rm bb}$,
was fixed to the value indicated in each panel. }
\label{fig:Sa_ga_kt}
\end{figure*}

In Figure \ref{fig:Sa_ga_kt} we show the asymptotic power-law photon index, $\Gamma$, 
as a function of $S_{\rm a}$ in 4U 1636--53. The value of $kT_{\rm bb}$  indicated in each panel 
was fixed during the fittings. When the source evolves from the hard spectral 
state to the transitional state in the CD, the value of $S_{\rm a}$ increases from 1.0 
to 2.0, $\Gamma$ increases from 1.8 to 2.2 and the source spectrum becomes
soft. As the source moves through the vertex in the CD, $S_{\rm a} \sim 2.1$,  $\Gamma$ 
drops sharply to $\sim 1.6$, and covers the range from $\sim 1.6$ to $\sim 2.2$. 
When the source moves out of the vertex to the 
bottom right part of the CD $\Gamma$ increases to $\sim 2.2$, and finally decreases back 
to $1.8-2.0$ as $S_{\rm a}$ increases to $\sim$2.5. It is clear from all panels that 
$\Gamma$ shows a significant drop at $S_{\rm a} \sim 2.1$, regardless of the value 
we choose for $kT_{\rm bb}$. Moreover, at low $S_{\rm a}$ values ($S_{\rm a} < 2.0$), 
$\Gamma$ shows a larger spread for low $kT_{\rm bb}$ values. This likely indicates that 
the lower blackbody temperature (e.g. $< 1.6$ keV) yields comparatively less 
well-constrained fits of the thermal Comptonisation model in 4U 1636$-$53.

\begin{figure*}
\begin{center}
\includegraphics[scale=0.39,angle=0]{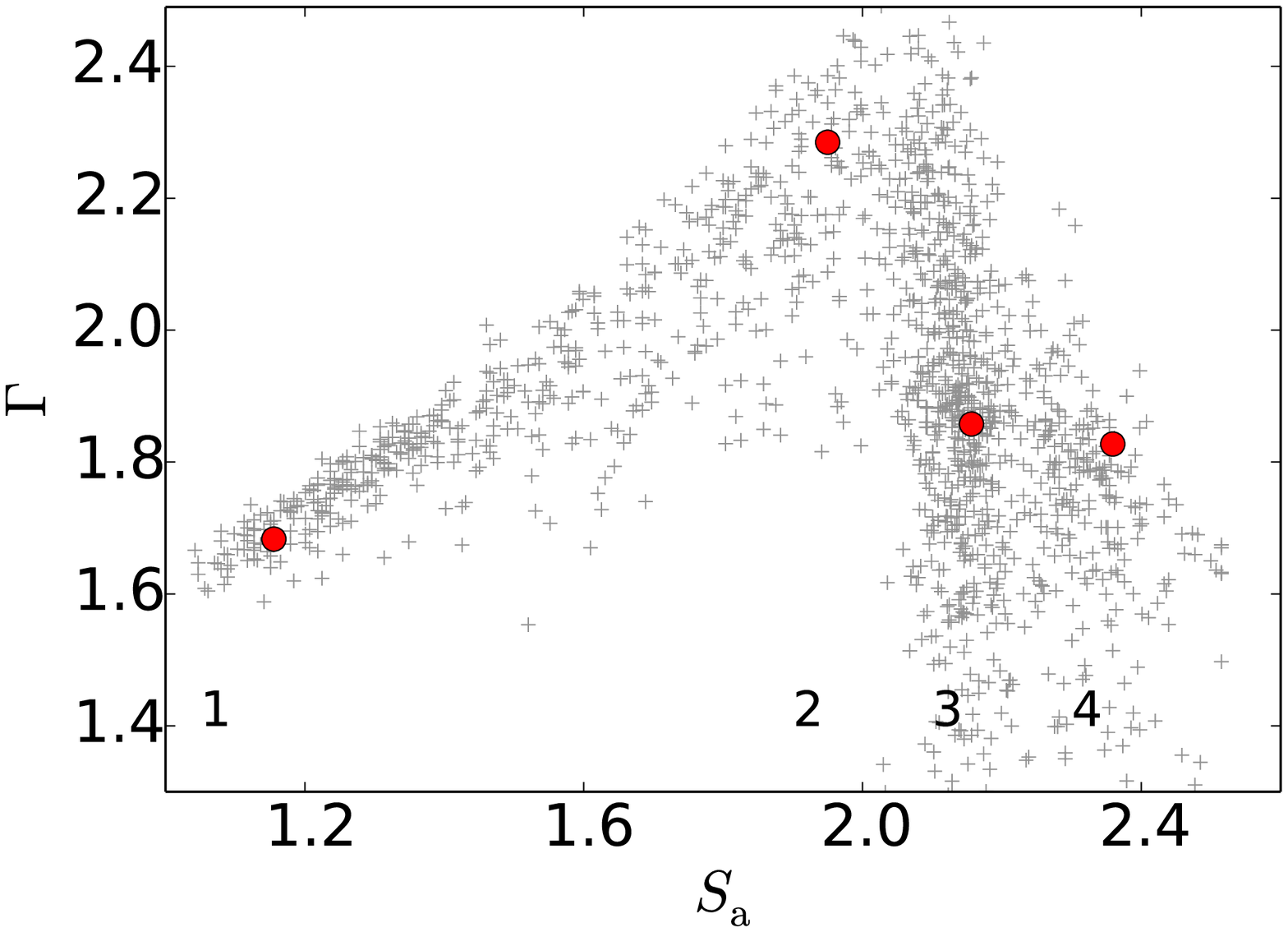}  
\includegraphics[scale=0.39,angle=0]{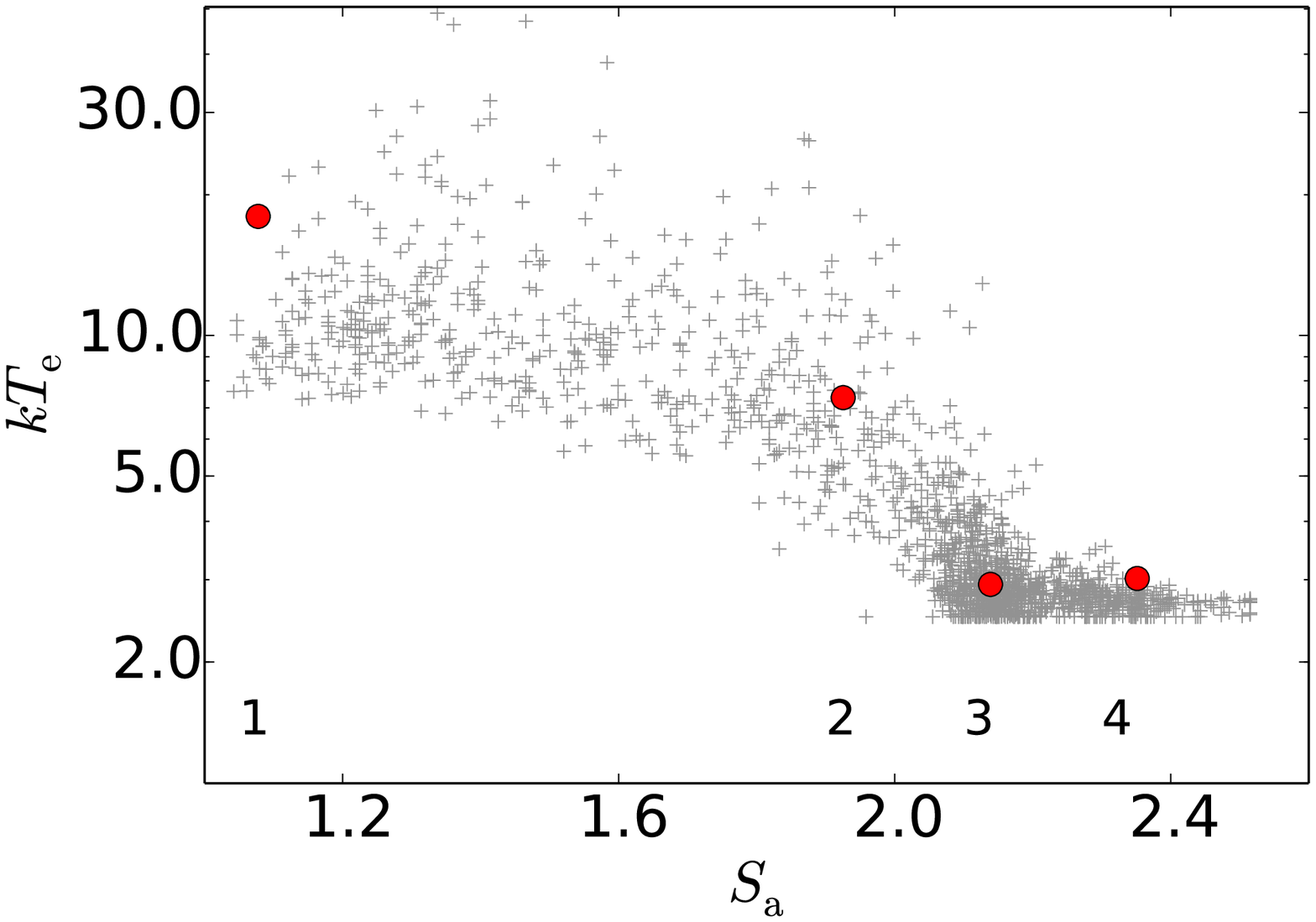}  \\
\includegraphics[scale=0.31,angle=270]{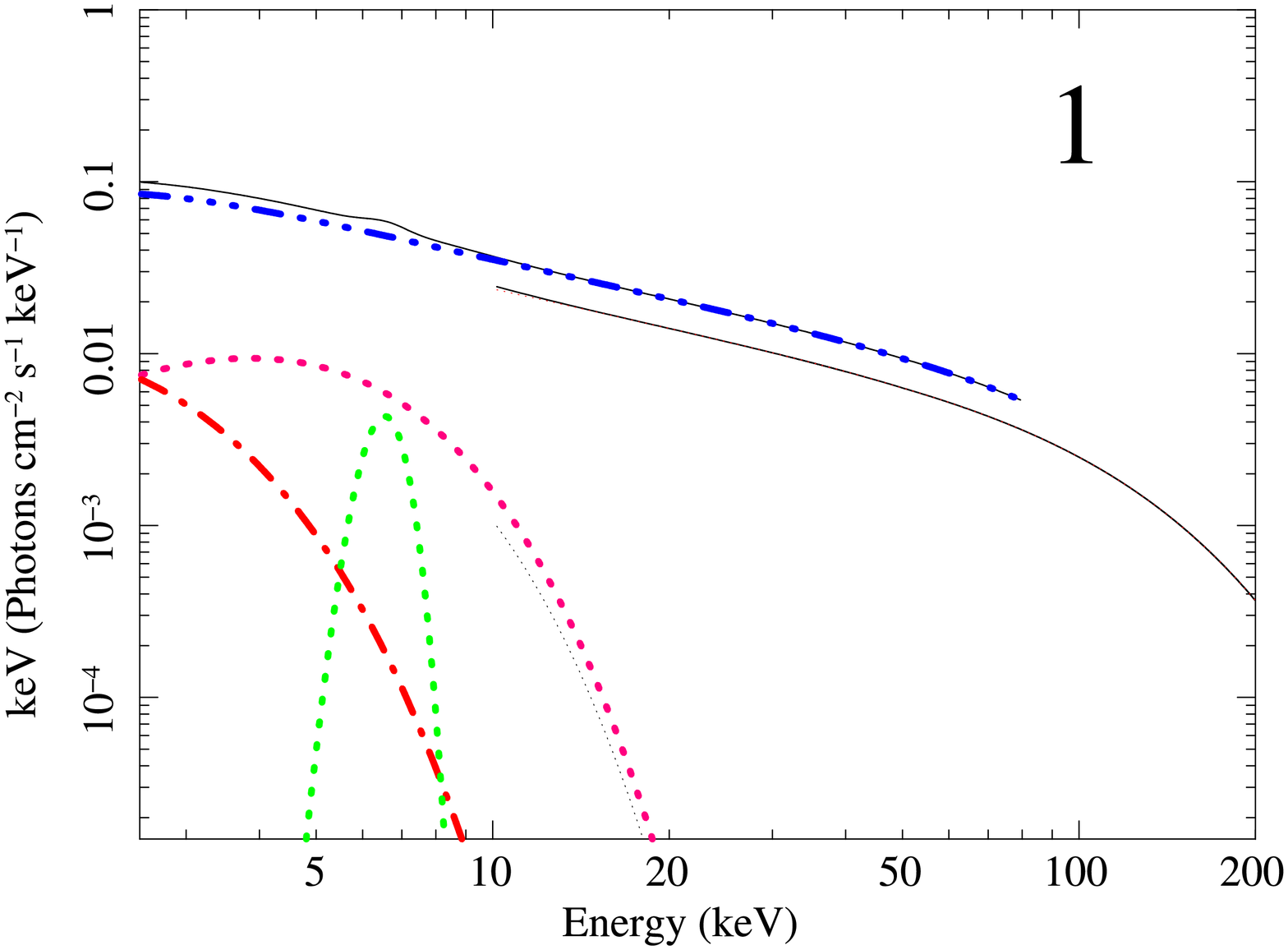}  
\includegraphics[scale=0.31,angle=270]{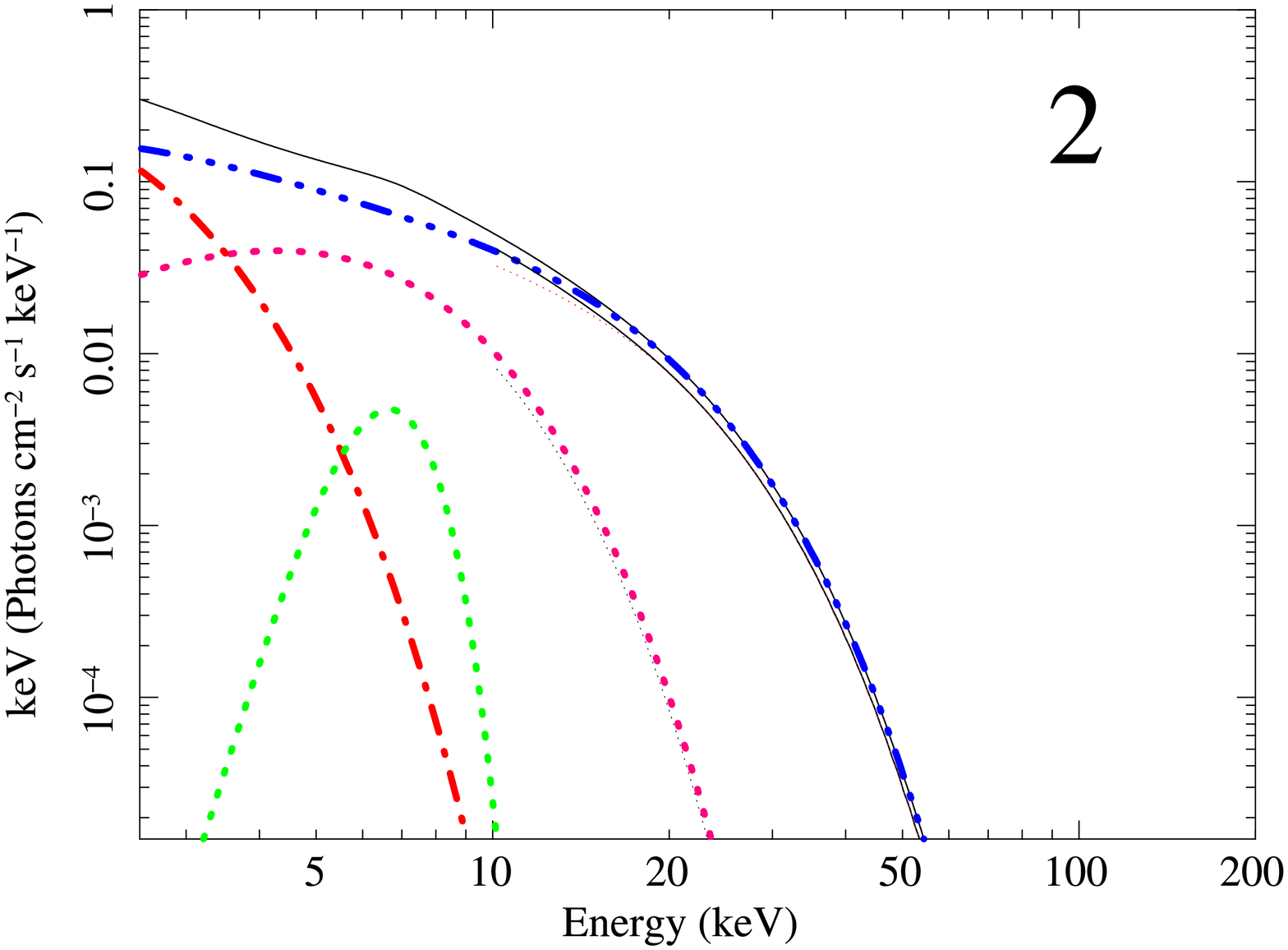} \\
\includegraphics[scale=0.31,angle=270]{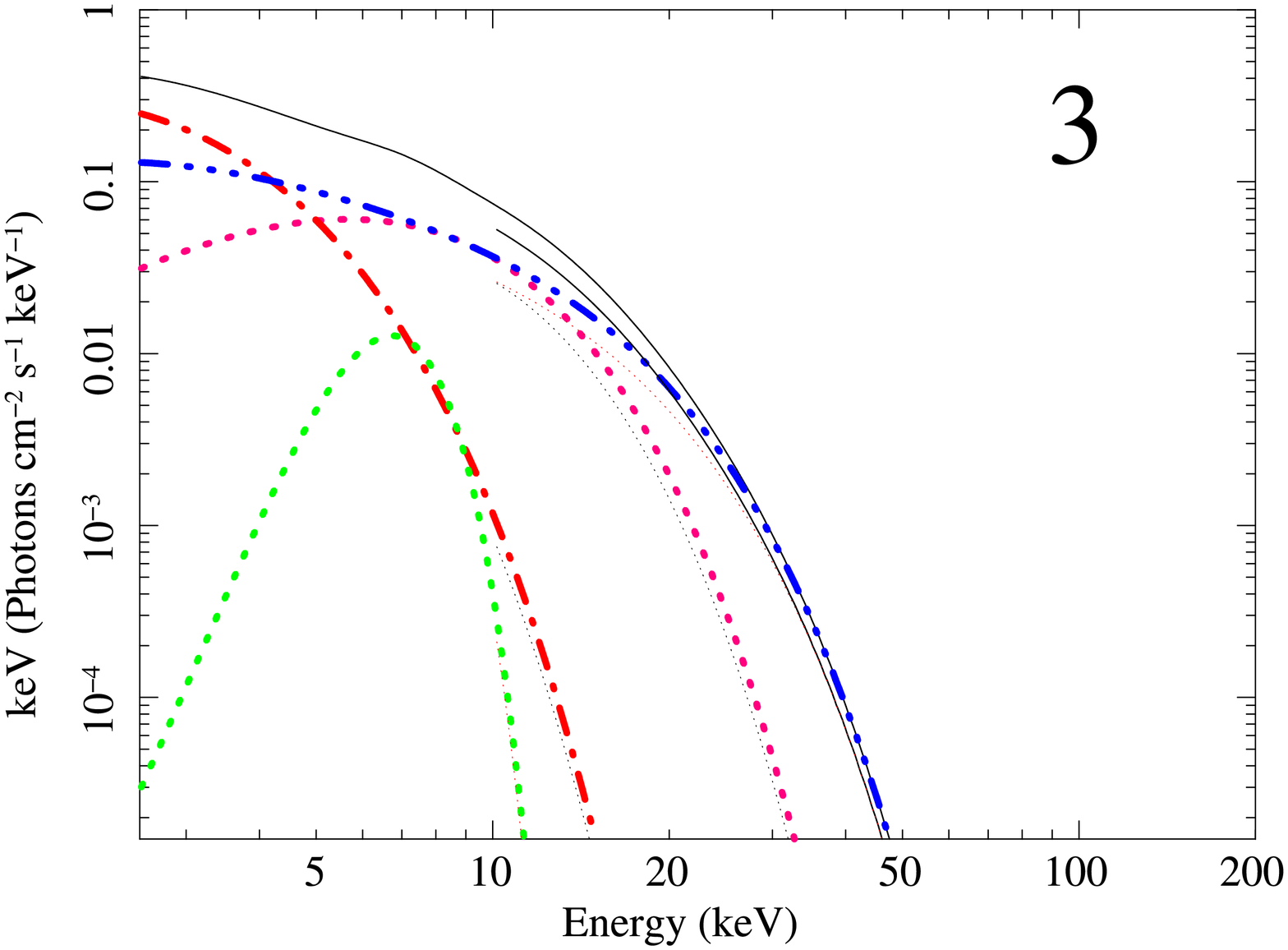}  
\includegraphics[scale=0.31,angle=270]{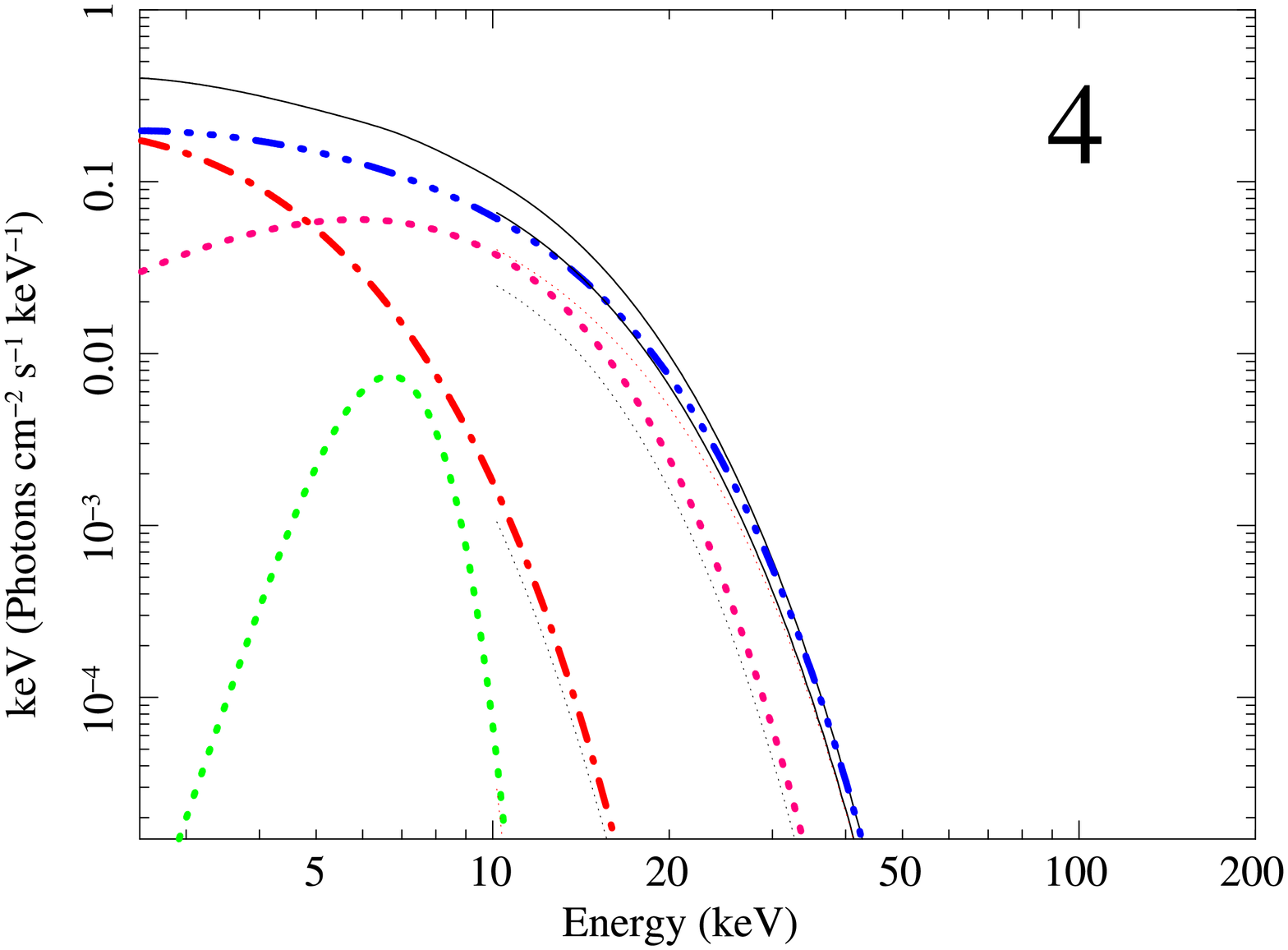}
\end{center}
\caption{
The upper left and right panels show, respectively, $\Gamma$ and $kT_{\rm e}$ of {\sc nthcomp} 
as a function of $S_{\rm a}$ in 4U 1636--53. Each gray cross corresponds to an 
individual observation. The red circles indicate four observations at different 
positions in the CD. The middle and lower panels show, respectively, the total 
model spectrum and the individual spectral components of these four observations 
(\textit{middle-left}, Obs1, \textit{middle-right}, Obs2, 
\textit{bottom-left}, Obs3, and \textit{bottom-right}, Obs4).  
The spectral components in the plots are {\sc diskbb} (red/dashed-dotted), 
{\sc bbodyrad}  (pink/dotted), {\sc gaussian} (green/dotted) and {\sc nthcomp} 
(blue dashed-three dotted), respectively. In each panel the solid black line 
represents the total model spectrum.}  
\label{fig:Sa_ga_pha}
\end{figure*}

The upper left panel of Figure \ref{fig:Sa_ga_pha} shows $\Gamma$ 
as a function of $S_{\rm a}$ when we used the interpolated values of $kT_{\rm dbb}$ and left
$kT_{\rm bb}$ free. The trend in this Figure is similar to those in Figure \ref{fig:Sa_ga_kt}. 
In the upper right panel Figure \ref{fig:Sa_ga_pha} we show the electron temperature, 
$kT_{\rm e}$, as a function  of $S_{\rm a}$.  As the source evolves 
from the hard to the transitional, and then to the soft spectral state
$kT_{\rm e}$ first decreases from 
$\sim 30$ keV to $\sim 3$ keV and then stays more or less constant.

In order to visualise the evolution of the individual components and the total spectrum 
in different states, we chose four observations covering the whole range of 
$S_{\rm a}$ values. The four observations, Obs1, Obs2, Obs3 and Obs4 with  $S_{\rm a} = $1.05, 
1.90, 2.15 and 2.35, respectively, are shown with red filled circles in the upper panels of 
Figure \ref{fig:Sa_ga_pha}.  In the middle and lower panels of Figure \ref{fig:Sa_ga_pha} 
we show the PCA/HEXTE model spectra of these 4 observations. 
The spectral components in the plots are {\sc diskbb} (red/dashed-dotted), 
{\sc bbodyrad}  (pink/dotted), {\sc gaussian} (green/dotted) and {\sc nthcomp} 
(blue dashed-three dotted), respectively. The best-fitting results for the 
four observations are given in Table \ref{tab-nthcomp}.

In the middle-left panel of Figure \ref{fig:Sa_ga_pha} (Obs1) the source is in 
the low/hard state.  The broad band X-ray spectrum is dominated by 
the hard/Comptonised component, {\sc nthcomp}. In this observation the power-law 
index is $\Gamma \sim 1.7$ and the electron temperature is $ kT_{\rm e} \sim 
17$ keV

When the source evolves from the low/hard state towards the vertex in the CD, the value 
of $S_{\rm a}$ increases from 1.0 to 2.0, $\Gamma$ increases from  $\sim 1.7$ to $\sim 2.3$ 
and the spectrum becomes soft (Obs2 in the right panel of Figure \ref{fig:Sa_ga_pha}).  
The electron temperature decreases from $kT_{\rm e} \sim 17$ keV  to $\sim 8$ keV. 
Compared to Obs1,  it is clear that the contribution of {\sc bbodyrad} and {\sc diskbb} 
increases, and the emission above 80 keV drops, significantly.

When the source is at the transitional state, $S_{\rm a} \sim 1.9 - 2.2$,  the value of 
$\Gamma$ covers a large range between 1.3 and 2.5. The emission from the {\sc diskbb} and 
{\sc nthcomp} components is more or less the same below 5 keV (bottom left panel of Figure \ref{fig:Sa_ga_pha}).
The {\sc nthcomp} component is relatively flat but the high-energy cut-off is at much lower 
energy than in Obs1. Our results are consistent with those of \cite{Sanna13} for the 
combined {\it XMM-Newton--RXTE} spectra (see their Figure 4).

When the source moves out from the transitional state to the high-soft state (Obs4),  
similar to what  \cite{Sanna13} show in their Figure 4, the normalization of {\sc diskbb} 
decreases and the emission from {\sc diskbb} is lower than that of {\sc nthcomp} below 
5 keV.

\begin{table}
\begin{center}
\tabcolsep=2.0mm
{
\caption{Best-fitting results for the four observations. The error bars are 
    given in 1 $\sigma$ level.}\label{tab-nthcomp}
\resizebox{1\linewidth}{!}{
\begin{tabular}{clcccccc}\hline

\multicolumn{1}{c}{Component} & 
\multicolumn{1}{l}{Parameter} & 
\multicolumn{1}{c}{Obs. 1} &  
\multicolumn{1}{c}{Obs. 2} &  
\multicolumn{1}{c}{Obs. 3} &  
\multicolumn{1}{c}{Obs. 4}\\\hline
\textsc{diskbb}  & $kT_{\rm dbb}$ (keV)         & 0.30                       & 0.60                       & 0.75                         & 0.80                  \\
                 & $N_{\rm dbb}$               & 550$\pm$320                & 120$\pm$70                 & 115$\pm$ 35                  & 165$\pm$85            \\
\hline
\textsc{bbody}   & $kT_{\rm bb}$ (keV)         & 1.33$\pm$0.12              & 1.75$\pm$0.28              & 2.05$\pm$0.27                & 1.72$\pm 0.65$        \\
		 & $N_{\rm bb}$ ($10^{-3}$)    & 1.3$\pm$0.8                & 3.8$\pm$2.4                & 2.9$^{+1.0}_{-0.4}$          & 6.2$^{+0.5}_{-0.2}$   \\
\hline
\textsc{nthcomp}& $\Gamma$                     & 1.68$\pm$0.05              & 2.28$\pm$0.27              & 1.86$\pm$0.39                & 1.81$\pm$0.42         \\
                 & $kT_{e}$ (keV)              & 17.1$\pm$6.1               & 8.2$\pm$4.7                & 2.9$\pm$0.4                  & 3.1$\pm$0.5 \\
                 & $N_{\rm NTH}$               & 0.17$\pm$0.02              & 0.12$\pm$0.04              & 0.31$\pm$0.09                & 0.19$\pm$0.07         \\
\hline
\end{tabular}}
\normalsize}
\end{center}
\end{table}

\subsection{X-ray spectral properties in the transitional state}
\label{xray_spectral_transaction}

\begin{figure}
    \centering
        \includegraphics[width=3.30in,angle=0]{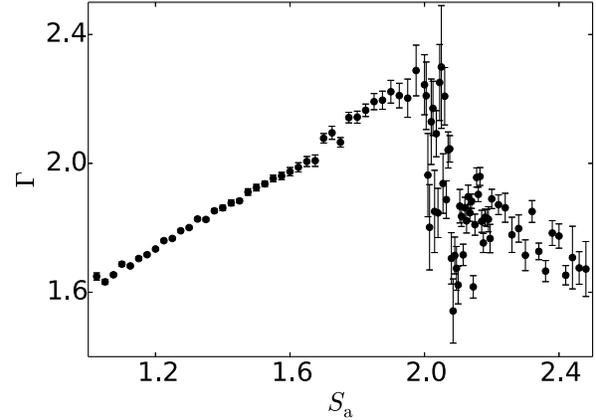}
    \caption{The power-law index, $\Gamma$, in {\sc nthcomp} as 
    a function of $S_{\rm a}$ in 4U 1636$-$53. The data are the same as in the upper 
    left panel of Figure \ref{fig:Sa_ga_pha}, but have been rebinned as described 
    in the text.   
             }
    \label{fig:Sa_gamma_rebin}
\end{figure}

From Figure \ref{fig:Sa_ga_pha} and the discussion above, 
it is apparent that the power-law photon index, $\Gamma$, in many of the observations 
shows a significant drop when the source is in the transitional state. 
Due to the large spread of $\Gamma$ when 2 $\simless$ $S_{\rm a}$ $\simless$ 2.2 
the trend of $\Gamma$ in this area of the plot is not clear.
We therefore first sorted the data in Figure \ref{fig:Sa_ga_pha} according 
to $S_{\rm a}$ and rebinned them using a step of 0.025, 0.005 and 0.02 in the range of 1.0$<S_{\rm a}<$1.8, 
1.8$<S_{\rm a}<$2.2 and $S_{\rm a}>$2.2, respectively. The weighted average $\Gamma$ as 
a function of $S_{\rm a}$ is shown in Figure \ref{fig:Sa_gamma_rebin}. From this Figure 
it is apparent that,  as $S_{\rm a}$ increases from $\sim2.0$ to $\sim2.2$,  $\Gamma$ 
decreases abruptly from $\sim2.3$ to $\sim1.5$, around $S_{\rm a}$  $\sim2.2$ $\Gamma$ 
increases again with $S_{\rm a}$ from $\sim1.5$ to $\sim1.9$ 
and, as $S_{\rm a}$ increases from $\sim2.2$ to $\sim2.5$,  $\Gamma$ decreases from 
$\sim1.9$ to $\sim1.7$. In the transitional state $\Gamma$ shows a large spread, 
but the plot in Figure \ref{fig:Sa_gamma_rebin} shows that $\Gamma$ changes significantly
with $S_{\rm a}$ and that the variation of $\Gamma$ is larger than 
the statistical fluctuations in the data. The large range of $\Gamma$ 
at the transitional state might be due to the fact that the model assumptions 
(e.g., a geometrically thin disc, or a spherical corona) are no longer valid.

\subsection{Relation between $\Gamma$ and the presence of the lower kHz QPO}
\label{timing and spectrum}

\begin{figure}
    \centering
        \includegraphics[width=3.30in,angle=0]{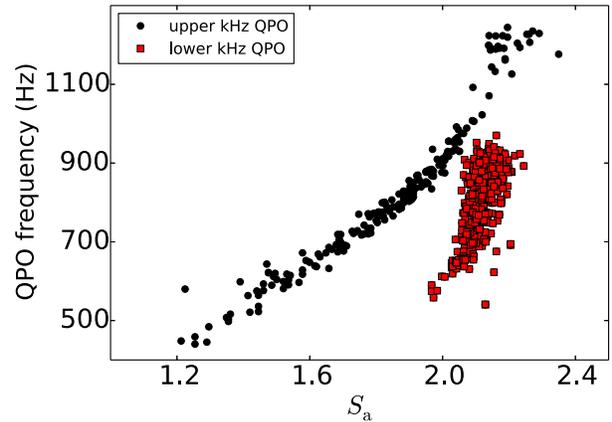}
    \caption{The kHz QPO frequency as a function of $S_{\rm a}$ in 4U 1636--53. 
    The filled circles (black) and filled squares (red) represent the 
    observations with upper and lower kHz QPOs, respectively.
             }
    \label{fig:Sa_qpo}
\end{figure}

Figure \ref{fig:Sa_qpo} shows the centroid frequency of all the kHz QPOs detected
in 4U 1636--53 as a function of $S_{\rm a}$.  The filled circles (black) and the 
filled squares (red) represent the observations with upper and lower kHz QPOs, 
respectively \citep[see also ][]{Sanna13}. The upper kHz
QPO is detected when $S_{\rm a} < 2.3$, and its frequency increases as 
$S_{\rm a}$ increases. The lower kHz QPO however, is only 
detected when $ 1.9 \simless S_{\rm a} \simless 2.2$, and its frequency covers a 
broad range over a narrow $S_{\rm a}$ range.

\begin{figure}
    \centering
        \includegraphics[width=3.30in,angle=0]{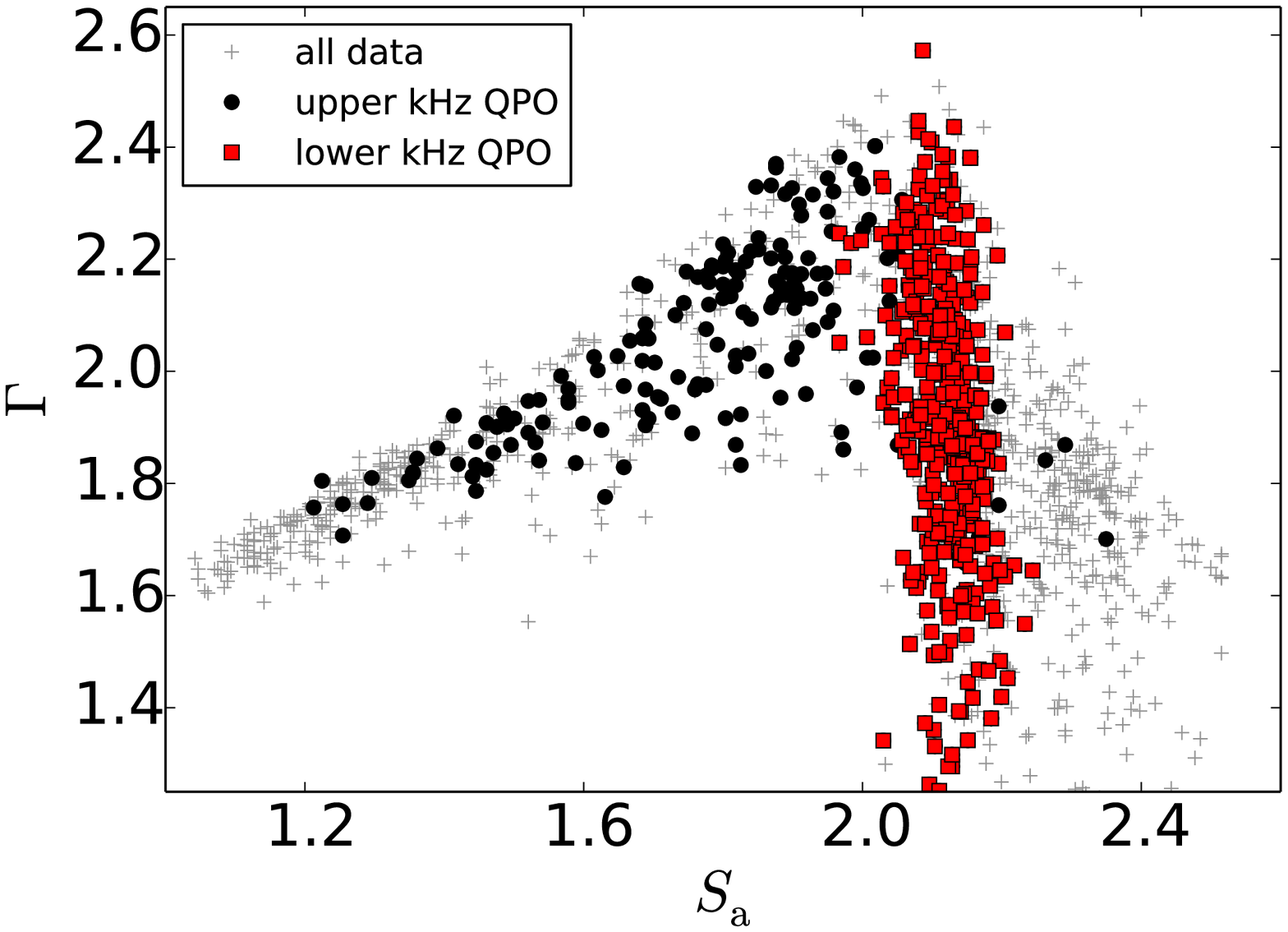}
        \includegraphics[width=3.30in,angle=0]{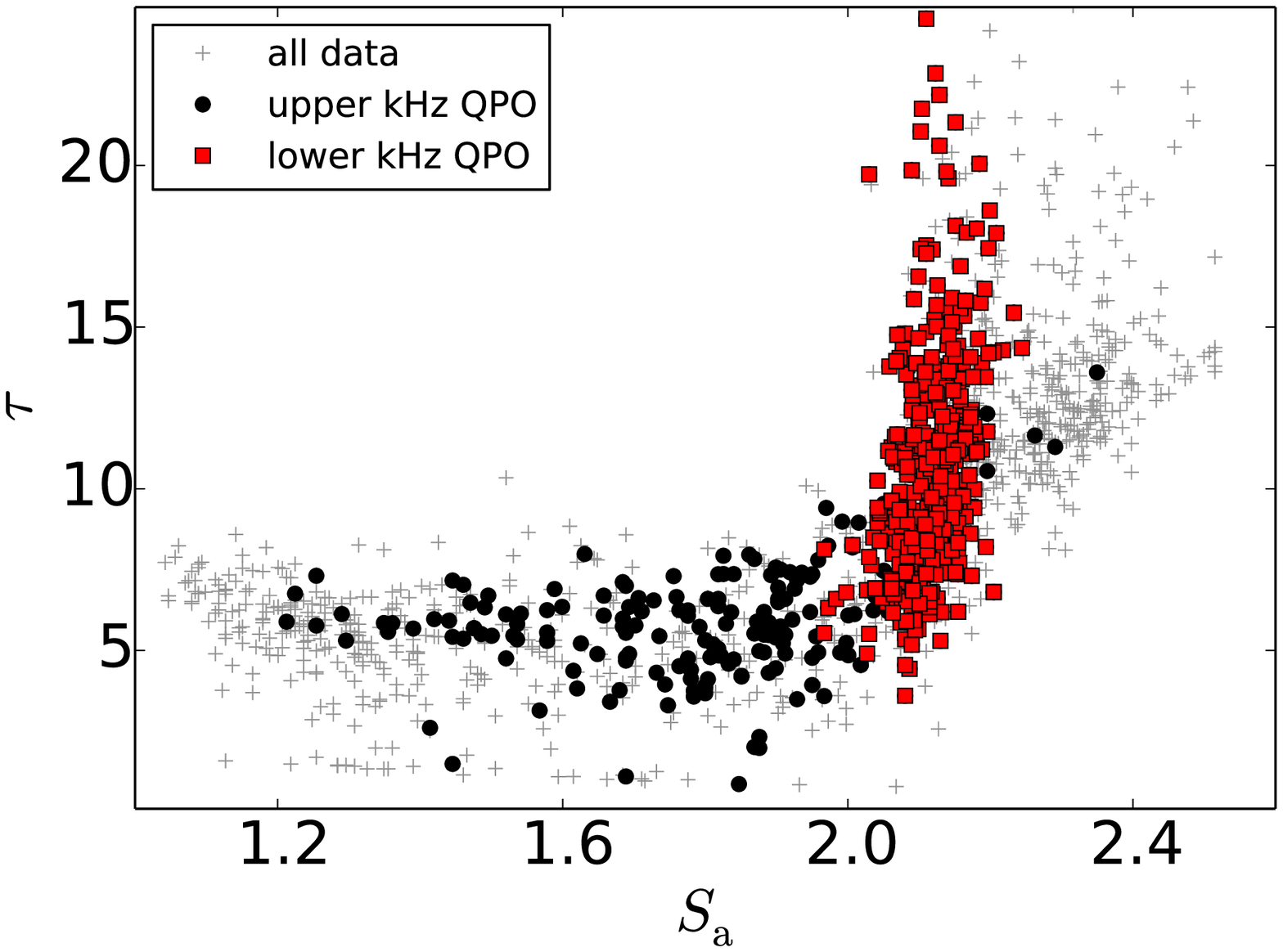}
    \caption{The upper and lower panels show, respectively, $\Gamma$ and $\tau$ as a 
    function of $S_{\rm a}$ in 4U 1636--53.
    The filled circles (black) and filled squares (red) represent the 
    observations with upper and lower kHz QPOs, respectively. The gray crosses 
    represent observations in which no kHz QPO was detected.
             }
    \label{fig:Sa_qpo_sigma}
\end{figure}

In the upper panel of Figure \ref{fig:Sa_qpo_sigma} we plot the power-law photon 
index, $\Gamma$, as a function of $S_{\rm a}$ (the same data shown in Figure 
\ref{fig:Sa_ga_pha}). The filled circles (black) and the filled squares (red) 
represent the  observations with upper and lower kHz QPOs, respectively.
Interestingly, observations with lower kHz QPOs correspond with those observations
in which $\Gamma$ shows a large spread.

In the case of Comptonisation in spherically symmetric region $\Gamma$ can be expressed as:

\begin{equation}
  \Gamma = \left[\frac{9}{4} + \frac{1}{(kT_e / m_e c^2) \tau (1 + \tau / 3)}\right]^{1/2} - \frac{1}{2}, 
  \label{eq_gamma}
\end{equation}

where $\tau$ is the optical depth of the medium, and $kT_e$ is the electron temperature 
of the Comptonising region \citep{Sunyaev80}. We calculated $\tau$ for each observation 
and plotted them as a function of $S_{\rm a}$ in the lower panel of Figure \ref{fig:Sa_qpo_sigma}. 
From this figure it is apparent that in the low/hard state the optical depth remains more 
or less constant at $\tau \sim 5$ as $S_{\rm a}$ increases from $\sim 1.0$ to $\sim 2.0$. 
When the source enters the transitional state, with $2.0 \simless S_{\rm a} \simless 2.2$, 
$\tau$ increases very abruptly with $S_{\rm a}$.
In the observations with a lower kHz QPO $\tau$ covers the range from $\sim 5$ to $\sim 25$.
A large value of $\tau$ has also been found in the soft state of other NS-LMXBs 
\citep[e.g. 4U 1608$-$52][]{GierlinskiDone02}. The large variation of $\Gamma$ 
(and $\tau$) over a small region in the CD indicates that, in the transitional 
state, either the spectrum changes significantly or the model is no longer 
physically appropriate. Even if the latter was the case, it is still interesting 
that the lower kHz QPO is detected only in this part of the diagram.

\section{Discussion}
\label{discussion}

We fitted the $3-180$ keV X-ray spectra of all {\it RXTE} observations of the neutron-star 
low-mass X-ray binary 4U 1636--53 with a model that includes a Comptonised component. 
We subsequently studied the relation between the parameters of this component and the 
presence of kHz QPOs as a function of the source state, using the parameter $S_{\rm a}$ 
to measure the position of the source in the CD. We found that, 
while the upper kHz QPO is present in almost all states of the source, the lower kHz QPO 
appears only during the transition from the hard to the soft state, in observations 
in which the optical depth of the Comptonised component is relatively high and increases 
very  sharply with $S_{\rm a}$, from $\tau\sim5$ to $\tau\sim25$.

In the low-hard state, the power-law index of the Comptonised component, $\Gamma$, 
increases gradually from $\Gamma\sim1.6$ to  $\Gamma \sim 2.3$ as the source spectrum 
softens and the source moves in the CD from the hard to the transitional state. 
In these observations $\Gamma$ is well correlated with the source hard colour and 
$S_{\rm a}$. The power density spectra of these observations show only the upper kHz 
QPO, with the QPO frequency being also correlated both with hard colour and $S_{\rm a}$. 
Similar correlations have been reported in other NS-LMXBs \citep[e.g., 4U 0614+09, 
4U 1608--52, 4U 1728--34, and Aql X--1; ][]{Kaaret98, Mendez99, Mendez99b, Mendez00}.

These correlations between spectral and timing properties in 4U 1636--53 are consistent 
with the scenario in which the truncation radius of the accretion disc decreases as mass 
accretion increases: As the X-ray luminosity of the source (and the inferred mass 
accretion rate through the disc) increases, the inner truncation radius of the accretion 
disc decreases \citep[e.g.][]{Done07,Mendez99,Mendez01, Straaten02,Altamirano08} and the 
disc temperature \citep{Shakura73}, and hence the soft flux, in the 
system increases. This in turn boosts the cooling of the corona through inverse Compton 
scattering, which leads to a drop of the electron temperature and an increase of the 
optical depth of the corona \citep{GierlinskiDone02}, and the energy spectrum of the 
source softens. In the case of 4U 1636--53 this picture is further supported by 
fits to the {\it XMM-Newton} and {\it RXTE} 
(PCA$+$HEXTE) spectra in the $0.8-200$ keV band over a range of source spectral states 
\citep{Sanna13, Lyu14}. Finally, if the frequency of the upper kHz QPO reflects the 
Keplerian frequency at the inner edge of the accretion disc, a smaller truncation 
radius yields a higher QPO frequency.

The above description only links the dynamics (frequency) of (at least) the upper kHz QPO 
to the spectral properties of the source, but it does not address the mechanism that 
determines the emission properties of the QPOs. For example, the steep increase of the 
rms amplitude of the kHz QPOs with energy \citep[e.g., ][]{Berger96, Mendez01, Gilfanov03} 
implies a modulation of the emitted flux of up to $\sim20$\% at $\sim25-30$ keV, 
where the contribution of the disc is negligible (see Figure \ref{fig:Sa_ga_pha}). 
Our findings suggest that the mechanisms that lead to the emission of the lower and the 
upper kHz QPO are not the same. This is similar to what was proposed by 
\cite{Marcio13} based on the energy and frequency dependence of the time 
lags of the kHz QPOs in this source.  We detect the upper kHz QPO over all source spectral 
states, except the extreme soft state,  whereas we only detect the lower kHz QPO in the 
transition from the hard to the soft state, at $S_{\rm a}\sim 1.9-2.2$,  where $\Gamma$ 
drops abruptly from $\sim 2.3$ to $\sim1.5$.

The parameter $\Gamma$, together with the electron temperature, $kT_e$, in the {\sc nthcomp} 
spectral model component is related to $\tau$, the optical depth of the corona responsible 
for the Comptonised emission, through eq.~(\ref{eq_gamma}). Since during the observations 
in the transitional state, where the lower kHz QPO is present, $\Gamma$ drops abruptly 
(see, e.g., the upper left panel of Figure~\ref{fig:Sa_ga_pha}), whereas $kT_{e}$ 
continues decreasing gradually (see the upper right panel of Figure~\ref{fig:Sa_ga_pha}), 
in the context of the {\sc nthcomp} model the drastic changes of $\Gamma$ likely reflect 
substantial changes of the optical depth of the corona. The lower panel of 
Figure~\ref{fig:Sa_qpo_sigma} shows that in the observations with the lower kHz QPO 
$\tau$ is high and increases very sharply with $S_{\rm a}$, from $\tau\sim5$ to $\tau\sim25$. 
Notwithstanding that in several of these observations we also detected the upper kHz QPO, 
this QPO is also present in observations in which the optical depth of the corona is low 
in comparison, $\tau\simless5$, and does not change much with $S_{\rm a}$.

Notice, by the way, that the drop of $\Gamma$ in the transitional and soft states does 
not contradict the fact that at $S_{\rm a} \ge 2$ the spectrum of the source softens. 
Indeed, although in the transitional and soft states $\Gamma$ decreases to values that 
are similar to those in the hard state, suggesting an overall hardening of the spectrum, 
$kT_e$ is much higher in the hard than in the transitional and soft states, such that 
the $3-180$ keV spectrum of the source does in fact softens as $S_{\rm a}$ increases 
(cf. Figure~\ref{fig:Sa_ga_pha}).

\cite{Lee01} \citep[see also][]{Lee98} proposed a model in which a feed-back loop between 
the corona and the source of soft photons (the disc and/or the neutron-star surface) 
excites a global mode in the system, such that the temperature and electron density of 
the corona and the temperature of the source of soft photons oscillate coherently at a 
given frequency, giving rise to a QPO. This model can explain the rms and 
time-lag energy spectra of the lower kHz QPO \citep{Berger96, Vaughan97} in the NS 
LMXB 4U 1608--52. This scenario also provides a
framework to interpret our findings in terms of a coupling and an efficient energy exchange 
between the disc and the corona. Since in the observations in which the lower kHz QPO is 
present the changes of the temperature of the corona are relatively small (at least 
compared to the relative changes of the optical depth), the frequency range spanned 
by this QPO likely reflects the range of optical depths (electron density) in the 
corona at which the resonance is excited.

We define the optical depth of the corona as 
$\tau = \sigma_{\rm T}n_{\rm e}l$ \citep{Lee98}, where $l$ is the size of the corona,   
$\sigma_{\rm T} = 6.65 \times10^{-25}$ cm$^2$ is the Thomson cross section of the electron, 
and $n_{\rm e}$ is the electron density. 
If we take 5 and 20 as extreme values of $\tau$ in our fits (see Figure \ref{fig:Sa_qpo_sigma}),
then 
$n_{\rm e}  \approx$ $0.5-1.9\times10^{19}$ cm$^{-3}$
and $2-7.5\times10^{19}$ cm$^{-3}$ for $l$ in the range of 4 and 15 km \citep{Lee01}, 
respectively.   These values are consistent with the 
typical electron density in the corona of neutron star or black hole systems 
\citep[e.g., ][]{Nobili00}.

It is interesting that in 4U 1636--53 the frequency at which (i) the lower kHz QPO 
is most often detected \citep{Belloni05}, (ii) the quality 
factor ($Q$, defined as the QPO frequency divided by the full-width at half-maximum) 
of the lower kHz QPO is the highest \citep{Barret06}, 
and (iii) the coherence between the low- and high-energy signals is the highest 
\citep{Marcio13}, is always $\sim 850$ Hz. Furthermore, \cite{Marcio16} 
recently showed that the phase lag between the low- and high-energy 
photons in the lower kHz QPO in 4U 1636--53 is largest at $S_{\rm a}\sim2.1$, 
which also corresponds to a QPO frequency of $\sim 850$ Hz.

All the above suggests that this frequency corresponds to a global oscillation 
mode of the accretion flow in this source \citep[see ][]{Lee01}. The mode could 
possibly be excited over a fairly large range of values of the properties of the 
corona (e.g., $\tau$, $kT_e$ and $l$), and hence over a range of frequencies of 
the QPO, as observed. On the other hand, the excitation and damping mechanisms 
of this global mode of the corona will play a role upon the range of frequencies 
and the coherence of the QPO. In a follow-up paper (Ribeiro et al. 2017, 
in preparation) we  discuss correlations between the rms amplitude of 
the QPOs and the spectral properties of the source, which provide further 
support to this scenario.

\section*{Acknowledgements}
This research has made use of data obtained from the High Energy
Astrophysics Science Archive Research Center (HEASARC), provided 
by NASA's Goddard Space Flight Center and NASA's Astrophysics Data 
System Bibliographic Services. We thank Tomaso Belloni for 
useful comments and discussions and Wenfei Yu for carefully reading 
and commenting the manuscript. ER acknowledges the support 
from Conselho Nacional de Desenvolvimento Científico e Tecnológico 
(CNPq - Brazil)


\begin{thebibliography}{}

\bibitem[\protect\citeauthoryear{{Altamirano}, {Ingram}, {van der Klis},
  {Wijnands}, {Linares} \& {Homan}}{{Altamirano} et~al.}{2012}]{Altamirano12}
{Altamirano} D.,  {Ingram} A.,  {van der Klis} M.,  {Wijnands} R.,  {Linares}
  M.,    {Homan} J.,  2012, \apjl, 759, L20

\bibitem[\protect\citeauthoryear{{Altamirano}, {van der Klis}, {M{\'e}ndez},
  {Jonker}, {Klein-Wolt} \& {Lewin}}{{Altamirano} et~al.}{2008}]{Altamirano08}
{Altamirano} D.,  {van der Klis} M.,  {M{\'e}ndez} M.,  {Jonker} P.~G.,
  {Klein-Wolt} M.,    {Lewin} W.~H.~G.,  2008, \apj, 685, 436

\bibitem[\protect\citeauthoryear{{Anders} \& {Grevesse}}{{Anders} \&
  {Grevesse}}{1989}]{Anders89}
{Anders} E.,  {Grevesse} N.,  1989, \gca, 53, 197

\bibitem[\protect\citeauthoryear{{Arnaud}}{{Arnaud}}{1996}]{Arnaud96}
{Arnaud} K.~A.,  1996, in {G.~H.~Jacoby \& J.~Barnes} ed., Astronomical Data
  Analysis Software and Systems V Vol.~101 of Astronomical Society of the
  Pacific Conference Series, {XSPEC: The First Ten Years}.
p.~17

\bibitem[\protect\citeauthoryear{{Balucinska-Church} \&
  {McCammon}}{{Balucinska-Church} \& {McCammon}}{1992}]{Balucinska92}
{Balucinska-Church} M.,  {McCammon} D.,  1992, \apj, 400, 699

\bibitem[\protect\citeauthoryear{{Barret}}{{Barret}}{2001}]{Barret01}
{Barret} D.,  2001, Advances in Space Research, 28, 307

\bibitem[\protect\citeauthoryear{{Barret}, {Olive} \& {Miller}}{{Barret}
  et~al.}{2006}]{Barret06}
{Barret} D.,  {Olive} J.-F.,    {Miller} M.~C.,  2006, \mnras, 370, 1140

\bibitem[\protect\citeauthoryear{{Belloni}, {Homan}, {Motta}, {Ratti} \&
  {M{\'e}ndez}}{{Belloni} et~al.}{2007}]{Belloni07}
{Belloni} T.,  {Homan} J.,  {Motta} S.,  {Ratti} E.,    {M{\'e}ndez} M.,  2007,
  \mnras, 379, 247

\bibitem[\protect\citeauthoryear{{Belloni}, {M{\'e}ndez} \& {Homan}}{{Belloni}
  et~al.}{2005}]{Belloni05}
{Belloni} T.,  {M{\'e}ndez} M.,    {Homan} J.,  2005, \aap, 437, 209

\bibitem[\protect\citeauthoryear{{Berger}, {van der Klis}, {van Paradijs},
  {Lewin}, {Lamb}, {Vaughan}, {Kuulkers}, {Augusteijn}, {Zhang}, {Marshall},
  {Swank}, {Lapidus}, {Lochner} \& {Strohmayer}}{{Berger}
  et~al.}{1996}]{Berger96}
{Berger} M.,  {van der Klis} M.,  {van Paradijs} J.,  {Lewin} W.~H.~G.,  {Lamb}
  F.,  {Vaughan} B.,  {Kuulkers} E.,  {Augusteijn} T.,  {Zhang} W.,  {Marshall}
  F.~E.,  {Swank} J.~H.,  {Lapidus} I.,  {Lochner} J.~C.,    {Strohmayer}
  T.~E.,  1996, \apjl, 469, L13

\bibitem[\protect\citeauthoryear{{de Avellar}, {M{\'e}ndez}, {Altamirano},
  {Sanna} \& {Zhang}}{{de Avellar} et~al.}{2016}]{Marcio16}
{de Avellar} M.~G.~B.,  {M{\'e}ndez} M.,  {Altamirano} D.,  {Sanna} A.,
  {Zhang} G.,  2016, \mnras, 461, 79

\bibitem[\protect\citeauthoryear{{de Avellar}, {M{\'e}ndez}, {Sanna} \&
  {Horvath}}{{de Avellar} et~al.}{2013}]{Marcio13}
{de Avellar} M.~G.~B.,  {M{\'e}ndez} M.,  {Sanna} A.,    {Horvath} J.~E.,
  2013, \mnras, 433, 3453

\bibitem[\protect\citeauthoryear{{Done}, {Gierli{\'n}ski} \& {Kubota}}{{Done}
  et~al.}{2007}]{Done07}
{Done} C.,  {Gierli{\'n}ski} M.,    {Kubota} A.,  2007, \aapr, 15, 1

\bibitem[\protect\citeauthoryear{{Gierli{\'n}ski} \& {Done}}{{Gierli{\'n}ski}
  \& {Done}}{2002}]{GierlinskiDone02}
{Gierli{\'n}ski} M.,  {Done} C.,  2002, \mnras, 337, 1373

\bibitem[\protect\citeauthoryear{{Gilfanov}, {Revnivtsev} \&
  {Molkov}}{{Gilfanov} et~al.}{2003}]{Gilfanov03}
{Gilfanov} M.,  {Revnivtsev} M.,    {Molkov} S.,  2003, \aap, 410, 217

\bibitem[\protect\citeauthoryear{{Hasinger} \& {van der Klis}}{{Hasinger} \&
  {van der Klis}}{1989}]{Hasinger89}
{Hasinger} G.,  {van der Klis} M.,  1989, \aap, 225, 79

\bibitem[\protect\citeauthoryear{{Jahoda}, {Markwardt}, {Radeva}, {Rots},
  {Stark}, {Swank}, {Strohmayer} \& {Zhang}}{{Jahoda} et~al.}{2006}]{Jahoda06}
{Jahoda} K.,  {Markwardt} C.~B.,  {Radeva} Y.,  {Rots} A.~H.,  {Stark} M.~J.,
  {Swank} J.~H.,  {Strohmayer} T.~E.,    {Zhang} W.,  2006, \apjs, 163, 401

\bibitem[\protect\citeauthoryear{{Jonker}, {M{\'e}ndez} \& {van der
  Klis}}{{Jonker} et~al.}{2002}]{Jonker02}
{Jonker} P.~G.,  {M{\'e}ndez} M.,    {van der Klis} M.,  2002, \mnras, 336, L1

\bibitem[\protect\citeauthoryear{{Kaaret}, {Yu}, {Ford} \& {Zhang}}{{Kaaret}
  et~al.}{1998}]{Kaaret98}
{Kaaret} P.,  {Yu} W.,  {Ford} E.~C.,    {Zhang} S.~N.,  1998, \apjl, 497, L93

\bibitem[\protect\citeauthoryear{{Kluzniak} \& {Abramowicz}}{{Kluzniak} \&
  {Abramowicz}}{2001}]{Kluzniak01}
{Kluzniak} W.,  {Abramowicz} M.~A.,  2001, Acta Physica Polonica B, 32, 3605

\bibitem[\protect\citeauthoryear{{Klu{\'z}niak}, {Abramowicz}, {Kato}, {Lee} \&
  {Stergioulas}}{{Klu{\'z}niak} et~al.}{2004}]{Kluzniak04}
{Klu{\'z}niak} W.,  {Abramowicz} M.~A.,  {Kato} S.,  {Lee} W.~H.,
  {Stergioulas} N.,  2004, \apjl, 603, L89

\bibitem[\protect\citeauthoryear{{Leahy}, {Darbro}, {Elsner}, {Weisskopf},
  {Kahn}, {Sutherland} \& {Grindlay}}{{Leahy} et~al.}{1983}]{Leahy83}
{Leahy} D.~A.,  {Darbro} W.,  {Elsner} R.~F.,  {Weisskopf} M.~C.,  {Kahn} S.,
  {Sutherland} P.~G.,    {Grindlay} J.~E.,  1983, \apj, 266, 160

\bibitem[\protect\citeauthoryear{{Lee} \& {Miller}}{{Lee} \&
  {Miller}}{1998}]{Lee98}
{Lee} H.~C.,  {Miller} G.~S.,  1998, \mnras, 299, 479

\bibitem[\protect\citeauthoryear{{Lee}, {Misra} \& {Taam}}{{Lee}
  et~al.}{2001}]{Lee01}
{Lee} H.~C.,  {Misra} R.,    {Taam} R.~E.,  2001, \apjl, 549, L229

\bibitem[\protect\citeauthoryear{{Lin}, {Remillard} \& {Homan}}{{Lin}
  et~al.}{2007}]{Lin07}
{Lin} D.,  {Remillard} R.~A.,    {Homan} J.,  2007, \apj, 667, 1073

\bibitem[\protect\citeauthoryear{{Lyu}, {M{\'e}ndez}, {Sanna}, {Homan},
  {Belloni} \& {Hiemstra}}{{Lyu} et~al.}{2014}]{Lyu14}
{Lyu} M.,  {M{\'e}ndez} M.,  {Sanna} A.,  {Homan} J.,  {Belloni} T.,
  {Hiemstra} B.,  2014, \mnras, 440, 1165

\bibitem[\protect\citeauthoryear{{M{\'e}ndez}}{{M{\'e}ndez}}{2000}]{Mendez00}
{M{\'e}ndez} M.,  2000, Nuclear Physics B Proceedings Supplements, 80, C1516

\bibitem[\protect\citeauthoryear{{M{\'e}ndez}}{{M{\'e}ndez}}{2006}]{Mendez06}
{M{\'e}ndez} M.,  2006, \mnras, 371, 1925

\bibitem[\protect\citeauthoryear{{M{\'e}ndez} \& {van der Klis}}{{M{\'e}ndez}
  \& {van der Klis}}{1999}]{Mendez99b}
{M{\'e}ndez} M.,  {van der Klis} M.,  1999, \apjl, 517, L51

\bibitem[\protect\citeauthoryear{{M{\'e}ndez}, {van der Klis} \&
  {Ford}}{{M{\'e}ndez} et~al.}{2001}]{Mendez01}
{M{\'e}ndez} M.,  {van der Klis} M.,    {Ford} E.~C.,  2001, \apj, 561, 1016

\bibitem[\protect\citeauthoryear{{M{\'e}ndez}, {van der Klis}, {Ford},
  {Wijnands} \& {van Paradijs}}{{M{\'e}ndez} et~al.}{1999}]{Mendez99}
{M{\'e}ndez} M.,  {van der Klis} M.,  {Ford} E.~C.,  {Wijnands} R.,    {van
  Paradijs} J.,  1999, \apjl, 511, L49

\bibitem[\protect\citeauthoryear{{Miller}, {Lamb} \& {Cook}}{{Miller}
  et~al.}{1998}]{Miller98}
{Miller} M.~C.,  {Lamb} F.~K.,    {Cook} G.~B.,  1998, \apj, 509, 793

\bibitem[\protect\citeauthoryear{{Nobili}, {Turolla}, {Zampieri} \&
  {Belloni}}{{Nobili} et~al.}{2000}]{Nobili00}
{Nobili} L.,  {Turolla} R.,  {Zampieri} L.,    {Belloni} T.,  2000, \apjl, 538,
  L137

\bibitem[\protect\citeauthoryear{{Plant}, {Fender}, {Ponti}, {Mu{\~n}oz-Darias}
  \& {Coriat}}{{Plant} et~al.}{2015}]{Plant15}
{Plant} D.~S.,  {Fender} R.~P.,  {Ponti} G.,  {Mu{\~n}oz-Darias} T.,
  {Coriat} M.,  2015, \aap, 573, A120

\bibitem[\protect\citeauthoryear{{Rothschild}, {Blanco}, {Gruber}, {Heindl},
  {MacDonald}, {Marsden}, {Pelling}, {Wayne} \& {Hink}}{{Rothschild}
  et~al.}{1998}]{Rothschild98}
{Rothschild} R.~E.,  {Blanco} P.~R.,  {Gruber} D.~E.,  {Heindl} W.~A.,
  {MacDonald} D.~R.,  {Marsden} D.~C.,  {Pelling} M.~R.,  {Wayne} L.~R.,
  {Hink} P.~L.,  1998, \apj, 496, 538

\bibitem[\protect\citeauthoryear{{Sanna}, {Hiemstra}, {M{\'e}ndez},
  {Altamirano}, {Belloni} \& {Linares}}{{Sanna} et~al.}{2013}]{Sanna13}
{Sanna} A.,  {Hiemstra} B.,  {M{\'e}ndez} M.,  {Altamirano} D.,  {Belloni} T.,
    {Linares} M.,  2013, \mnras, 432, 1144

\bibitem[\protect\citeauthoryear{{Sanna}, {M{\'e}ndez}, {Belloni} \&
  {Altamirano}}{{Sanna} et~al.}{2012}]{Sanna12}
{Sanna} A.,  {M{\'e}ndez} M.,  {Belloni} T.,    {Altamirano} D.,  2012, \mnras,
  424, 2936

\bibitem[\protect\citeauthoryear{{Shakura} \& {Sunyaev}}{{Shakura} \&
  {Sunyaev}}{1973}]{Shakura73}
{Shakura} N.~I.,  {Sunyaev} R.~A.,  1973, \aap, 24, 337

\bibitem[\protect\citeauthoryear{{Stella} \& {Vietri}}{{Stella} \&
  {Vietri}}{1998}]{Stella98}
{Stella} L.,  {Vietri} M.,  1998, \apjl, 492, L59

\bibitem[\protect\citeauthoryear{{Sunyaev} \& {Titarchuk}}{{Sunyaev} \&
  {Titarchuk}}{1980}]{Sunyaev80}
{Sunyaev} R.~A.,  {Titarchuk} L.~G.,  1980, \aap, 86, 121

\bibitem[\protect\citeauthoryear{{van der Klis}}{{van der Klis}}{2006}]{Klis06}
{van der Klis} M.,  2006, {Rapid X-ray Variability}.
pp 39--112

\bibitem[\protect\citeauthoryear{{van Straaten}, {van der Klis}, {di Salvo} \&
  {Belloni}}{{van Straaten} et~al.}{2002}]{Straaten02}
{van Straaten} S.,  {van der Klis} M.,  {di Salvo} T.,    {Belloni} T.,  2002,
  \apj, 568, 912

\bibitem[\protect\citeauthoryear{{Vaughan}, {van der Klis}, {M{\'e}ndez}, {van
  Paradijs}, {Wijnands}, {Lewin}, {Lamb}, {Psaltis}, {Kuulkers} \&
  {Oosterbroek}}{{Vaughan} et~al.}{1997}]{Vaughan97}
{Vaughan} B.~A.,  {van der Klis} M.,  {M{\'e}ndez} M.,  {van Paradijs} J.,
  {Wijnands} R.~A.~D.,  {Lewin} W.~H.~G.,  {Lamb} F.~K.,  {Psaltis} D.,
  {Kuulkers} E.,    {Oosterbroek} T.,  1997, \apjl, 483, L115

\bibitem[\protect\citeauthoryear{{Wijnands}, {van der Klis}, {van Paradijs},
  {Lewin}, {Lamb}, {Vaughan} \& {Kuulkers}}{{Wijnands}
  et~al.}{1997}]{Wijnands97}
{Wijnands} R.~A.~D.,  {van der Klis} M.,  {van Paradijs} J.,  {Lewin} W.~H.~G.,
   {Lamb} F.~K.,  {Vaughan} B.,    {Kuulkers} E.,  1997, \apjl, 479, L141

\bibitem[\protect\citeauthoryear{{Zdziarski}, {Johnson} \&
  {Magdziarz}}{{Zdziarski} et~al.}{1996}]{Zdziarski1996}
{Zdziarski} A.~A.,  {Johnson} W.~N.,    {Magdziarz} P.,  1996, \mnras, 283, 193

\bibitem[\protect\citeauthoryear{{Zhang}, {M{\'e}ndez}, {Altamirano}, {Belloni}
  \& {Homan}}{{Zhang} et~al.}{2009}]{Zhanggb09}
{Zhang} G.,  {M{\'e}ndez} M.,  {Altamirano} D.,  {Belloni} T.~M.,    {Homan}
  J.,  2009, \mnras, 398, 368

\bibitem[\protect\citeauthoryear{{Zhang}, {Lapidus}, {Swank}, {White} \&
  {Titarchuk}}{{Zhang} et~al.}{1997}]{Zhang97}
{Zhang} W.,  {Lapidus} I.,  {Swank} J.~H.,  {White} N.~E.,    {Titarchuk} L.,
  1997, \iaucirc, 6541

\bibitem[\protect\citeauthoryear{{{\.Z}ycki}, {Done} \& {Smith}}{{{\.Z}ycki}
  et~al.}{1999}]{Zycki1999}
{{\.Z}ycki} P.~T.,  {Done} C.,    {Smith} D.~A.,  1999, \mnras, 309, 561

\end{thebibliography}


\end{document}